\documentclass[aps,pra,eqsecnum,showpacs,superscriptaddress,twocolumn]{revtex4}
\usepackage{amsmath}
\usepackage{amssymb}

\usepackage{graphicx}
\usepackage{dcolumn}
\usepackage{bm} 
\usepackage{color}
\usepackage{epstopdf}

\def\bra#1{\langle #1 |}
\def\ket#1{| #1 \rangle}

\def\cH{\mathcal{H}}

\def\cS{\mathcal{S}}

\def\e{\mathrm{e}}
\def\i{\mathrm{i}}
\def\d{\mathrm{d}}

\def\tr{\mathop{\textrm{tr}}\nolimits}

\newcommand{\comm}[1]{}

\begin{document}
\title{A large-$N$ approximated field theory for multipartite entanglement}

\author{P. Facchi}
\affiliation{Dipartimento di Fisica and MECENAS, Universit\`a di Bari, I-70126 Bari, Italy} 
\affiliation{INFN, Sezione di Bari, I-70126 Bari, Italy}

\author{G. Florio}
\affiliation{Dipartimento di Fisica and MECENAS, Universit\`a di Bari, I-70126 Bari, Italy} 
\affiliation{INFN, Sezione di Bari, I-70126 Bari, Italy}
\affiliation{Museo Storico della Fisica e Centro Studi e Ricerche
``Enrico Fermi'', Piazza del Viminale 1, I-00184 Roma, Italy}
\affiliation{Dipartimento di Meccanica, Matematica e Management, Politecnico di Bari, Via E. Orabona 4, 70125 Bari, Italy}

\author{G. Parisi}
\affiliation{Dipartimento di Fisica, Universit\`{a} di Roma ``Sapienza'', Piazzale Aldo Moro 2, I-00185 Roma, Italy}
\affiliation{Centre for Statistical Mechanics and Complexity (SMC), CNR-INFM, I-00185 Roma, Italy}
\affiliation{INFN, Sezione di Roma,  I-00185 Roma, Italy}

\author{S. Pascazio}
\affiliation{Dipartimento di Fisica and MECENAS, Universit\`a di Bari, I-70126 Bari, Italy} 
\affiliation{INFN, Sezione di Bari, I-70126 Bari, Italy}

\author{A. Scardicchio}
\affiliation{International Center for Theoretical Physics, Strada Costiera 11, 34151 Trieste, Italy}
\affiliation{INFN, Sezione di Trieste, I-34151, Trieste, Italy}

\date{\today}
\begin{abstract}
We study the characterization of multipartite entanglement for the random states of an $n$-qbit system. Unable to solve the problem exactly we generalize it, changing complex numbers into real vectors with $N_c$ components (the original problem is recovered for $N_c=2$). Studying the leading diagrams in the large-$N_c$ approximation, we unearth the presence of a phase transition and, in an explicit example, show that the so-called entanglement frustration disappears in the large-$N_c$ limit.
\end{abstract}

\maketitle

\section{Introduction}\label{sec:intro}

The study of entanglement is almost as old as quantum mechanics, as it was the subject of seminal papers by Einstein, Podolsky and Rosen \cite{EPR} and Schr\"odinger \cite{Schr}. The focus of the founding fathers was on the puzzling, non-classical aspects of quantum correlations.
Nowadays, entanglement \cite{woot,fazio,horo} is viewed mostly as a crucial resource in quantum applications, quantum communication and quantum information processing \cite{nielsen_chuang}, and it mantains its original fascination for the comprehension of the structure and geometry of quantum mechanics \cite{BZ}.

While bipartite entanglement is well understood and quantified, the notion of multipartite entanglement is more elusive. 
This is due to a number of concomitant factors. First of all, for many-body quantum systems, the number of entanglement measures grows exponentially with the system size {\cite{mw}}, making a characterization of quantum correlations complicated \cite{MMSZ}.
Second, new properties {arise} when more quantum parties are involved, among these the intriguing appearance of frustration. In agreement with the classical notion \cite{MPV}, this is related to the impossibility of satisfying a number of requirements at the same time \cite{frustration}. Applied to entanglement, this means that given three (or more) parties A, B and C, if the entanglement between A and B grows, that between A and C or B and C decreases. This is also referred to as ``monogamy" of entanglement~\cite{ckw,verstraete,adesso}.

In our work the symptoms of such frustration will surface in the investigation of $n=4$ qubits~\cite{gourwallach,HS00} and for small number of qudits \cite{KNM,BSSB,OS,BPBZCP,BH07}, and cannot be avoided for $n\ge 8$ qubits, where one can prove that it is impossible to maximize the bipartite entanglement by maximizing indipendently all possible bipartitions of a system of qubits \cite{scott,mmes,facchi_lincei}. Further interesting contributions to these problems were given in Refs.\ \cite{AC,zycz}.

In this article we will study the properties of multipartite entanglement by adopting the concepts and tools of classical statistical mechanics \cite{FFP,multipartite,classical,depasquale3}. This approach has proved useful in the study of the bipartite entanglement of a large number of qubits, where it unearths the presence of phase transitions \cite{Facchi2,depasquale2}. The situation with multipartite entanglement is, however, much more complex. This is due to the fact that the monogamy of  entanglement acts effectively as a frustration, and the resulting statistical system is frustrated, to say, similarly to a spin glass.
 
In order to explore the rich landscape that ensues, we shall make use of techniques that are based on the analysis of diagrams that naturally arise when one considers a high-temperature expansion of the distribution function of the measure of multipartite entanglement (the potential of multipartite entanglement) \cite{multipartite}.

Unfortunately, the evaluation of the contributions of different kinds of graph and their resummation is not a simple task. One would like to find  a strategy that can {select} a procedure to sum some particular families of diagrams in order to analyze the features of multipartite entanglement. In this article we will therefore introduce a new parameter, the ``colour'' index $N_c$ (the case of qubits is recovered when $N_c=2$). This appears in a natural way by generalizing of the structure of the measure used to characterize the multipartite entanglement. We will consider the limit of large $N_c$ and reveal the presence of a phase transition. Moreover, we will see an explicit example where the frustration of multipartite entanglement disappears if the value of $N_c$ is large enough. This simplification for large $N_c$ is common in statistical models and field theories and the behavior of the large-$N_c$ approximation usually captures some (but never all) the behavior of the finite, or even small-$N_c$ theories. 

This article is organized as follows.
In Sec.\ \ref{statmech} we introduce notation and familiarize with the statistical mechanics approach we will use.
In Sec.\ \ref{cumul} we evaluate the first relevant cumulants of the theory.
We restrict our attention to an interesting class of diagrams in Sec.~\ref{cactus}: this enables us to look at the limit of large number of colors. We investigate the behavior of the energy (representing entanglement) in Sec.\ \ref{sec:minimum}: this unveils the presence of a phase transition, that is studied in Sec. \ref{sec:numeric}.
The phase transition is further investigated in Sec.\ \ref{sec:hysteresis}, where we numerically show that no hysteresis appears. We conclude in Sec.\ \ref{sec:conclusions} with a few comments.

\section{Multipartite Entanglement and Statistical Mechanics}
\label{statmech}

\subsection{Potential of multipartite entanglement}
Let us consider an ensemble $S=\{1,2,\dots, n\}$ of $n$ qubits in the
Hilbert space $\mathcal{H}_S= (\mathbb{C}^2)^{\otimes n}$. In this article we will focus
on pure states 
\begin{equation}
|z\rangle = \sum_{k\in {\mathbb{Z}}_2^n} z_k |k\rangle , \quad z_k \in
\mathbb{C}, \quad \sum_{k\in {\mathbb{Z}}_2^n} {|z_k|}^2 =1,
\label{eq:genrandomx}
\end{equation}
where $z=(z_k)$, $k=(k_i)_{i\in S}$,  $k_i\in {\mathbb{Z}}_2=\{0,1\}$, and
\begin{equation}
\label{eq:ki}
\ket{k}=\bigotimes_{i\in S} \ket{k_i}_i, \qquad
\ket{k_i}_i \in \mathbb{C}^2, \qquad \bra{k_i} k_j \rangle= \delta_{ij}  .
\end{equation}

Consider a bipartition $(A,\bar{A})$ of the system, where $A \subset S$ is a
subset of $n_A$ qubits and $\bar{A}=S\backslash A$ its complement,
with $n_A+n_{\bar{A}}=n$. We set $n_A
\leq n_{\bar{A}}$ without loss of generality. The total Hilbert space
factorizes into
$\mathcal{H}_S=\mathcal{H}_A\otimes\mathcal{H}_{\bar{A}}$, with
$\mathcal{H}_A= \bigotimes_{i\in A} \mathbb{C}^2_i$, of dimensions
$N_A=2^{n_A}$ and $N_{\bar{A}}=2^{n_{\bar{A}}}$, respectively
($N_AN_{\bar{A}}=N$). 

The \emph{bipartite}
entanglement between the two subsets can be measured by the purity of the reduced state of subsystem $A$
\begin{equation}
\pi_{A}(z)=\mathrm{Tr} \rho_{A}^2, \quad \rho_{A}=\mathrm{Tr}_{\bar{A}} \ket{\psi} \bra{\psi},
\label{eq:puritydef}
\end{equation}
$\rm{Tr}_{\bar{A}}$ being the partial trace over $\bar{A}$. We notice
that $\pi_{A}=\pi_{\bar{A}}$ and
\begin{equation}
\label{eq:purityconstraint}
1/N_A\le\pi_{A}\le 1,
\end{equation}
where the lower and upper bounds are obtained for maximally entangled and separable states, respectively.

A natural extension of this measure to the multipartite scenario is realized considering the \emph{potential of multipartite
entanglement}, that is the average bipartite
entanglement between balanced bipartitions \cite{classical}:
\begin{eqnarray}
H(z) 
&=&  \left(\begin{array}{l}n
\\n_A\end{array}\!\!\right)^{-1}\sum_{|A|=n_A}\pi_{A}( z )\nonumber\\&=& \sum_{k,k',l,l' \in {\mathbb{Z}}_2^n} \Delta(k, k'; l, l'
)\, z_{k}\, z_{k'}\,
\bar{z}_{l}\, \bar{z}_{l'}\, ,
\label{eq:pimeDelta}
\end{eqnarray}
where $n_A=[n/2]$ (balanced bipartitions), {with $[x]$ being the integer part of $x$}. The \emph{coupling function} is \cite{facchi_lincei,classical}
\begin{equation}\label{eq:deltag}
\Delta(k,k';l,l'
)= g\big((k\oplus l) \vee (k' \oplus l'), (k
\oplus l') \vee (k'\oplus l)
\big),
\end{equation}
where
\begin{equation}
 g(a,b
 ) = \delta_{a\wedge b,\,0} \; \hat{g}(|a|,|b|
 ) 
\label{eq:gdef}
\end{equation}
and
\begin{eqnarray}
\hat{g}(s,t 
)=
\frac{1}{2}\left(\!\!\begin{array}{c}n \\{n_A}\end{array}\!\!\right)^{\!\!-1}  \left[ \left(\!\!\begin{array}{c}
     n-s-t    \\
      {n_A}-s
\end{array}\!\! \right) +\left(\!\!\begin{array}{c}
     n-s-t    \\
      {n_A}-t
\end{array}\!\! \right)\right] .\nonumber\\
\label{eq:hatg}
\end{eqnarray}
In Eqs.~(\ref{eq:deltag})-(\ref{eq:gdef}) we have defined $|a|=\sum_{i\in S} a_i$, $|b|=\sum_{i\in S} b_i$, $a\oplus b=(a_i + b_i\; \mathrm{mod}\; 2)_{i \in S}$ being the XOR operation,  $a\vee b=(a_i + b_i - a_i b_i) _{i \in S}$ the OR operation,  $a\wedge b=(a_i b_i)_{i\in S}$ the AND operation.
Due to its linear structure, $H(z)$ inherits the upper and lower bound of the purity $\pi_A(z)$ in~(\ref{eq:purityconstraint}):
\begin{equation}
1/N_A\le H(z) \le 1.
\end{equation}

We notice the following symmetries of the
coupling function: 
\begin{eqnarray}
\label{eq:symmetries}
\left\{\begin{array}{c}
      \Delta(k,k';l,l')=\Delta(k',k;l,l')    \\
      \Delta(k,k';l,l')=\Delta(l,l';k,k')    \\
      \Delta(k,k';l,l')=\Delta(k',k;l',l)
\end{array} \right. ,
\end{eqnarray}
which reflect the reality of $H(z)$ 
\begin{equation}
\overline{H(z)}=H(\overline{z})=H(z).
\end{equation}
Moreover, since 
\begin{equation}
\Delta(k\oplus m,k'\oplus m;l\oplus m,l'\oplus m)=\Delta(k,k';l,l'),
\end{equation} 
the Hamiltonian is  invariant under rotations and reflections:
\begin{equation}
H (z_{k\oplus m} )= H (z_{k} ).
\end{equation}
for every $m\in {\mathbb{Z}}_2^n$.

\subsection{Classical statistical mechanics approach}

The analysis of the properties of $H(z)$ can
be rephrased in a classical statistical mechanical framework. 
Let us consider the partition function of a system with Hamiltonian $H(z)$ at a fictitious temperature $\beta^{-1}$,
\begin{eqnarray}
\label{partition.function}
Z(\beta) &=& \int \e^{-\beta H(z)}\; \d\mu(z), 
\end{eqnarray}
where \cite{Zyc2}
\begin{equation}
\d\mu(z)= \frac{(N-1)! }{\pi^N}\; \delta\!\left(1-\|z\|^2\right) \,
\prod_k {\d z_k \d\bar{z}_k} ,
\label{eq:meastyp}
\end{equation}
is the uniform measure 
on the hypersphere
\begin{equation}
\label{eq:constraint} 
\|z\|^2=\sum_k |z_k|^2=1,
\end{equation}
and $\d z_k \d\bar{z}_k$ denotes the Lebesgue measure on $\mathbb{C}$.

The Lagrange multiplier $\beta$ fixes the average value of multipartite entanglement. In particular, $\beta = 0$ corresponds to the {uniform} sampling of random states. The limits $\beta\rightarrow\pm\infty$ select configurations that, respectively, minimize or maximize $H(z)$; the former case corresponds to maximally multipartite entangled states (MMES) \cite{mmes}, the latter to completely separable states. 

The average value of $H(z)$ (seen as the average energy of the system) at arbitrary $\beta$ can be
obtained as
\begin{eqnarray} \label{mean.energy}
\langle H\rangle_\beta = \frac{1}{Z(\beta)}\int 
H\, \e^{-\beta H}\; \d\mu=-\frac{\partial}{\partial\beta}\ln Z(\beta).
\end{eqnarray}
The properties of the distribution function of the potential of
multipartite entanglement can be analyzed by evaluating its cumulants. In particular,  the $m$-th cumulant of $H(z)$ reads
\begin{equation}
\kappa^{(m)}_\beta
={(-1)}^m \frac{\partial^{m}}{\partial {\beta}^m}
\ln Z(\beta)
={(-1)}^{m-1} \frac{\partial^{m-1}}{\partial {\beta}^{m-1}}
\langle H\rangle_\beta.
\end{equation}

\section{Diagrammatic evaluation of the cumulants}
\label{cumul}
In this section we will briefly review some properties of the high temperature
expansion of the distribution function of the potential of
multipartite entanglement. We remind that for $\beta =0$
one gets the unbiased random states. 
The average energy 
reads
\begin{eqnarray}
\label{high.temp} \langle H\rangle_\beta
= \sum_{m=1}^\infty\frac{(-\beta)^{m-1}}{(m-1)!}\kappa_{0}^{(m)}
\label{eq:energyhightemp}
\end{eqnarray} 
where the brackets  $\langle \cdots \rangle_0$ denote
the average with respect to the {uniform (unitarily invariant)} measure (\ref{eq:meastyp}).
The only nonvanishing correlation functions are of the form~\cite{classical}
\begin{eqnarray} 
\left\langle \prod_{j=1}^{N} |z_{j}|^{2m_j} \right\rangle_0
&=&\frac{(N-1)!\,\prod_{j=1}^{N} m_j!}{\left(N-1+\sum_{j=1}^{N}
m_j\right)!}\;  ,
\label{exp}
\end{eqnarray}
where $m_j$ are nonnegative integers.
In order to calculate the required cumulants, we will make use of the diagrammatic technique developed in \cite{classical}. 
In this way, we will be able to evaluate the contribution arising from different kinds of graphs. The objective of this study is to try to understand whether by embedding this problem in a larger family, dependent on a parameter to be introduced in the following, called $N_c$, the number of colours, a class of graphs can be isolated that dominate in an appropriate limit. To this extent, we will promote the complex numbers defining the wave functions to real vectors of $N_c$ components and take $N_c$ large. In the following two sections we recall the computations in \cite{classical}, and we anticipate which diagrams are going to be relevant in the large $N_c$ limit.


\subsection{First cumulant}

\begin{figure}[t]
\begin{center}
\includegraphics[height=0.12\textwidth]{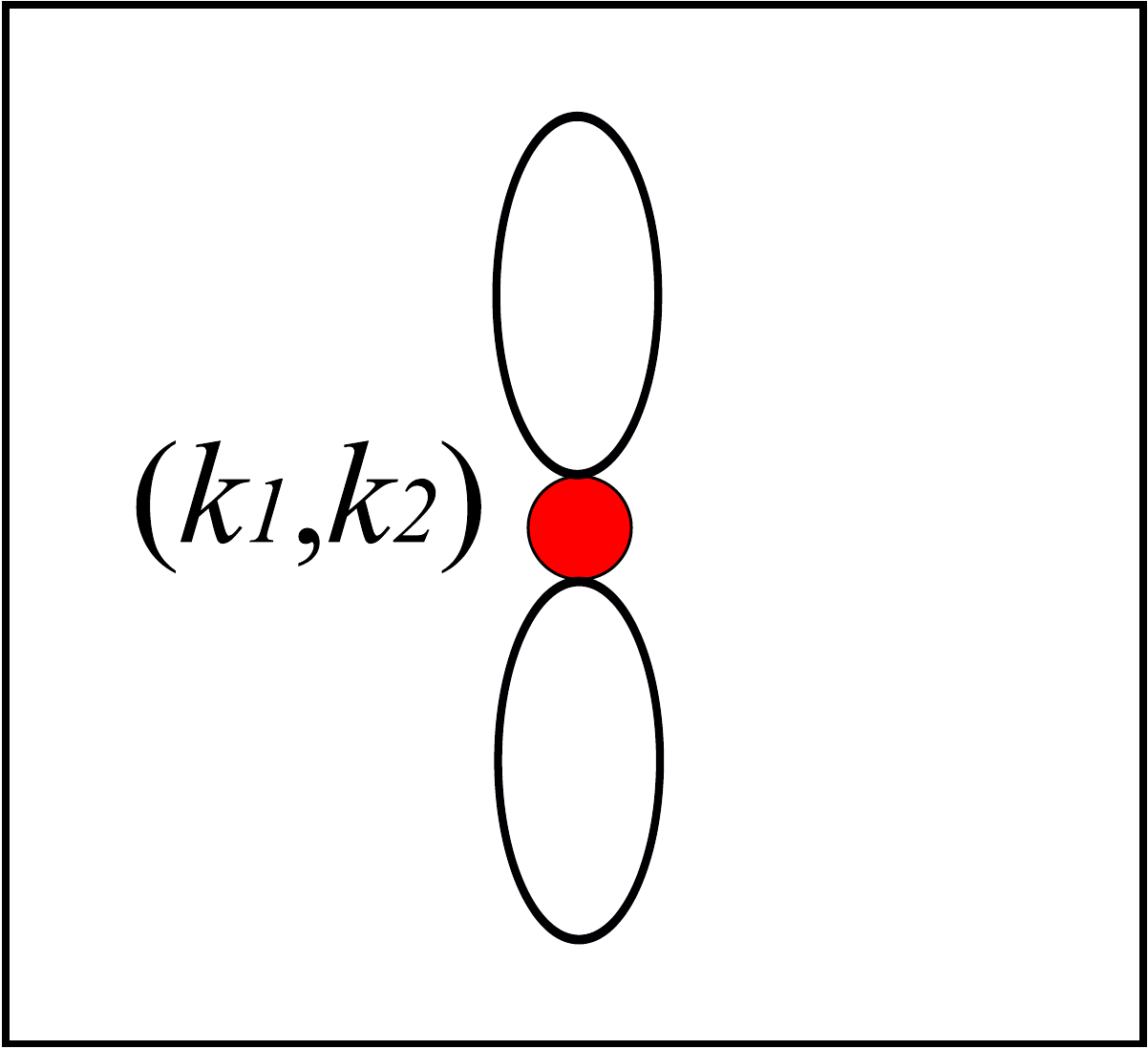}
\caption{(Color online) Graph contributing to the first cumulant: two-leaf cactus diagram.
}
\label{fig:graphfirstcumulant}
\end{center}
\end{figure}

The average energy at $\beta=0$ reads
\begin{eqnarray}
\langle H\rangle_0 &=& \sum_{k,l\in {\mathbb{Z}}_2^{2n}} \Delta (k_1,k_2;l_1,l_2)\langle
z_{k_1}z_{k_2}\bar{z}_{l_1}\bar{z}_{l_2}\rangle_0\nonumber\\
&=&  \langle |z_{1}|^2 |z_{2}|^2 \rangle_0 \sum_{p\in\mathcal{S}_2} [p(1)\; p(2)]
\end{eqnarray}
where
\begin{equation}
[p(1)\; p(2)]=  \sum_{k_1,k_2\in {\mathbb{Z}}_2^{n}} \Delta(k_1,k_2;k_{p(1)},k_{p(2)})
\end{equation}
and $\mathcal{S}_2$ is the symmetric group of order $2$.
This is represented by the doubly degenerate graph in  Fig.~\ref{fig:graphfirstcumulant} which, 
by~(\ref{eq:deltag}) and  (\ref{exp}) gives
\begin{equation}
\label{eq:firstcumul}
 \langle H\rangle_0 = \frac{N_A+N_{\bar{A}}}{N+1}.
\end{equation}
For balanced  bipartitions {of an even number of qubits},  $N_A=N_{\bar{A}}=\sqrt{N}$, and $N\rightarrow +\infty$ we get
\begin{equation}
\label{eq:firstcumul1}
 \langle H\rangle_0   =  \frac{2\sqrt{N}}{N+1} \sim \frac{2}{\sqrt{N}} .
\end{equation}
{For an odd number of qubits the value of $\langle H\rangle_0$  is $3/\sqrt{2}$ larger than~(\ref{eq:firstcumul1}).}

\subsection{Second cumulant}

The second cumulant is defined as
\begin{equation}
\label{eq:definitionk2}
\kappa_{0}^{(2)}
= \left\langle
H^2\right\rangle_0- \langle H\rangle_0^2.
\end{equation}
We have
\begin{eqnarray}
\langle H^2\rangle_0&=&\sum_{k,l\in {\mathbb{Z}}_2^{4n}} \Delta (k_1,k_2;l_1,l_2)\Delta (k_3,k_4;l_3,l_4) \nonumber\\
& & \times \langle
z_{k_1}z_{k_2}z_{k_3}z_{k_4}\bar{z}_{l_1}\bar{z}_{l_2}\bar{z}_{l_3}\bar{z}_{l_4}\rangle_0 \nonumber\\
&=&\langle |z_{1}|^{2} |z_{2}|^{2} |z_{3}|^{2} |z_{4}|^{2}\rangle_0\sum_{p\in\mathcal{S}_4}  [p(1)\; p(2), p(3)\; p(4)]. \nonumber\\
\end{eqnarray}

\begin{figure}
\begin{center}
\includegraphics[width=0.22\textwidth]{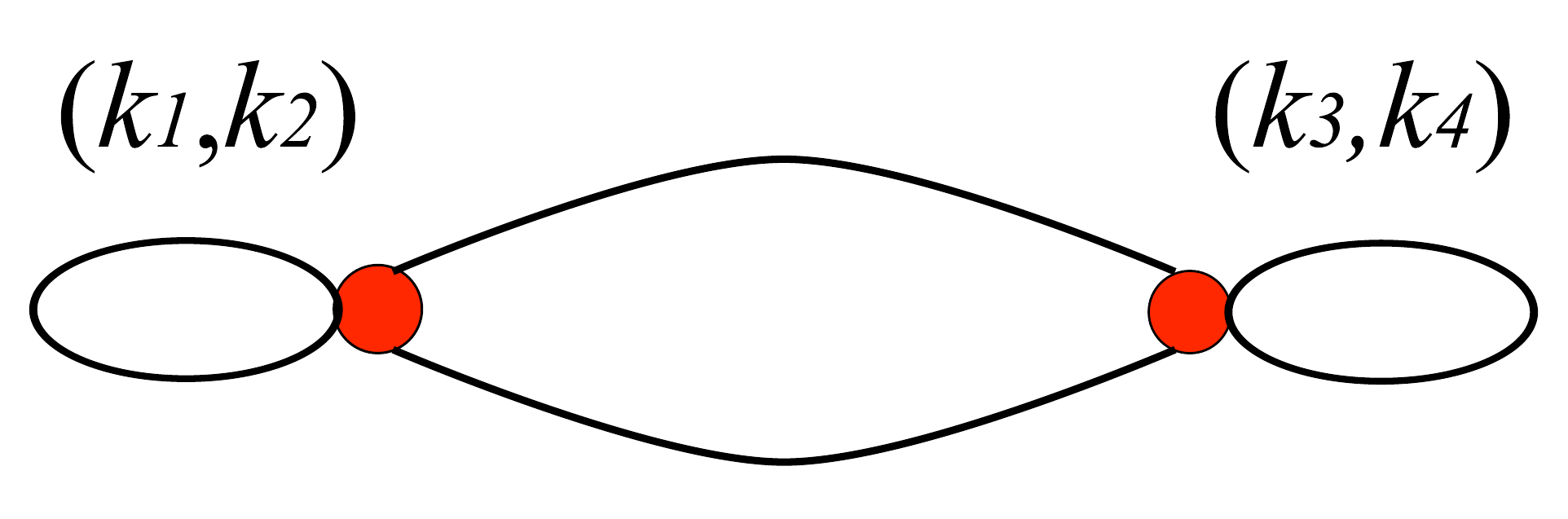}
\caption{(Color online) Connected graph contributing to the second cumulant: three-leaf cactus diagram.
}
\label{fig:graphsecondcum}
\end{center}
\end{figure}

\begin{figure}
\begin{center}
\includegraphics[width=0.22\textwidth]{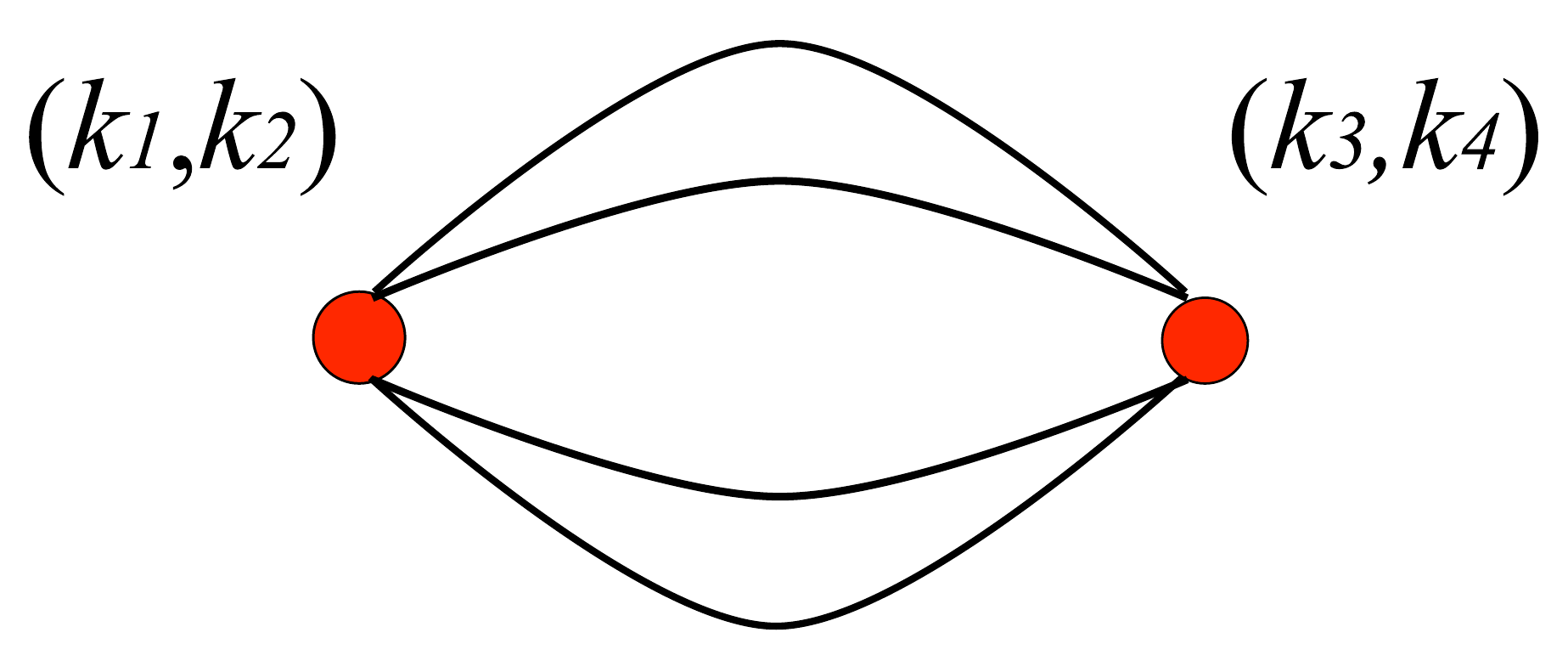}
\caption{(Color online) Connected graph without leaves contributing to the second cumulant.
}
\label{fig:graphsecondcumnoloop}
\end{center}
\end{figure}

Let us evaluate the contribution from connected graphs. The graph with two links between left and right pairs in Fig.~\ref{fig:graphsecondcum} has degeneracy $16$ and reads
\begin{eqnarray}\label{eq:graphseconcumul1}
[1\; 3, 2\; 4]\; &=& \frac{N\;{(N_A+N_{\bar{A}})}^2}{4}. 
\end{eqnarray}
The graph shown in Fig.~\ref{fig:graphsecondcumnoloop} has degeneracy $4$. The associated contribution does not have a transparent form; on the other hand, its asymptotic formula reads \cite{classical}
\begin{eqnarray}
[3\; 4, 1\; 2]\; \sim \frac{3\sqrt{2}}{4}N^{\alpha}
\label{eq:dominant2}
\end{eqnarray}
with $\alpha=\log_2 3\simeq1.5850$.

Notice the presence of the irrational exponent in the graph~(\ref{eq:dominant2}). Moreover, since~(\ref{eq:graphseconcumul1}) is exactly canceled by the non-connected contribution from the square of~(\ref{eq:firstcumul}),  the graph~(\ref{eq:dominant2}) represents the dominant contribution to the second cumulant $\kappa_0^{(2)}$ that therefore has the asymptotic value of
\begin{equation}
\kappa_0^{(2)}\simeq\frac{3\sqrt{2}}{N^{2.4150...}}.
\end{equation}
However, considering only the contribution from the graph~(\ref{eq:graphseconcumul1}), which will turn out to be dominant in the large $N_c$ limit (see later in the paper), we obtain
\begin{eqnarray}\label{eq:second}
\tilde{\kappa}_0^{(2)} &=&   
\frac{4 (N_A+N_{\bar{A}})^2}{(N+1)(N+2)(N+3)} \sim \frac{16}{N^2}
\end{eqnarray}
where the asymptotic expression is valid for balanced  bipartitions {of an even number of qubits},  $N_A=N_{\bar{A}}=\sqrt{N}$, and $N\rightarrow +\infty$.


\subsection{Third cumulant}
\label{sec:third_cumul}

The third cumulant reads
\begin{eqnarray}
\label{eq:definitionk3}
\kappa_{0}^{(3)} 
&=&\left\langle \left(
H- \langle H\rangle_0\right)^3\right\rangle_0\nonumber\\&=&\langle H^3\rangle_0-3\langle H^2\rangle_0\langle H\rangle_0+2\langle H\rangle_0^3,
\end{eqnarray}
and we have
\begin{eqnarray}
\langle H^3\rangle_0 &=& \langle |z_{1}|^{2} |z_{2}|^{2} |z_{3}|^{2} |z_{4}|^{2}|z_{5}|^{2} |z_{6}|^{2}\rangle_0\nonumber\\
&\times&\sum_{p\in\mathcal{S}_6}  [p(1)\; p(2), p(3)\; p(4), p(5)\; p(6)].\nonumber\\
\end{eqnarray}

\begin{figure}[t]
\begin{center}
\includegraphics[width=0.20\textwidth]{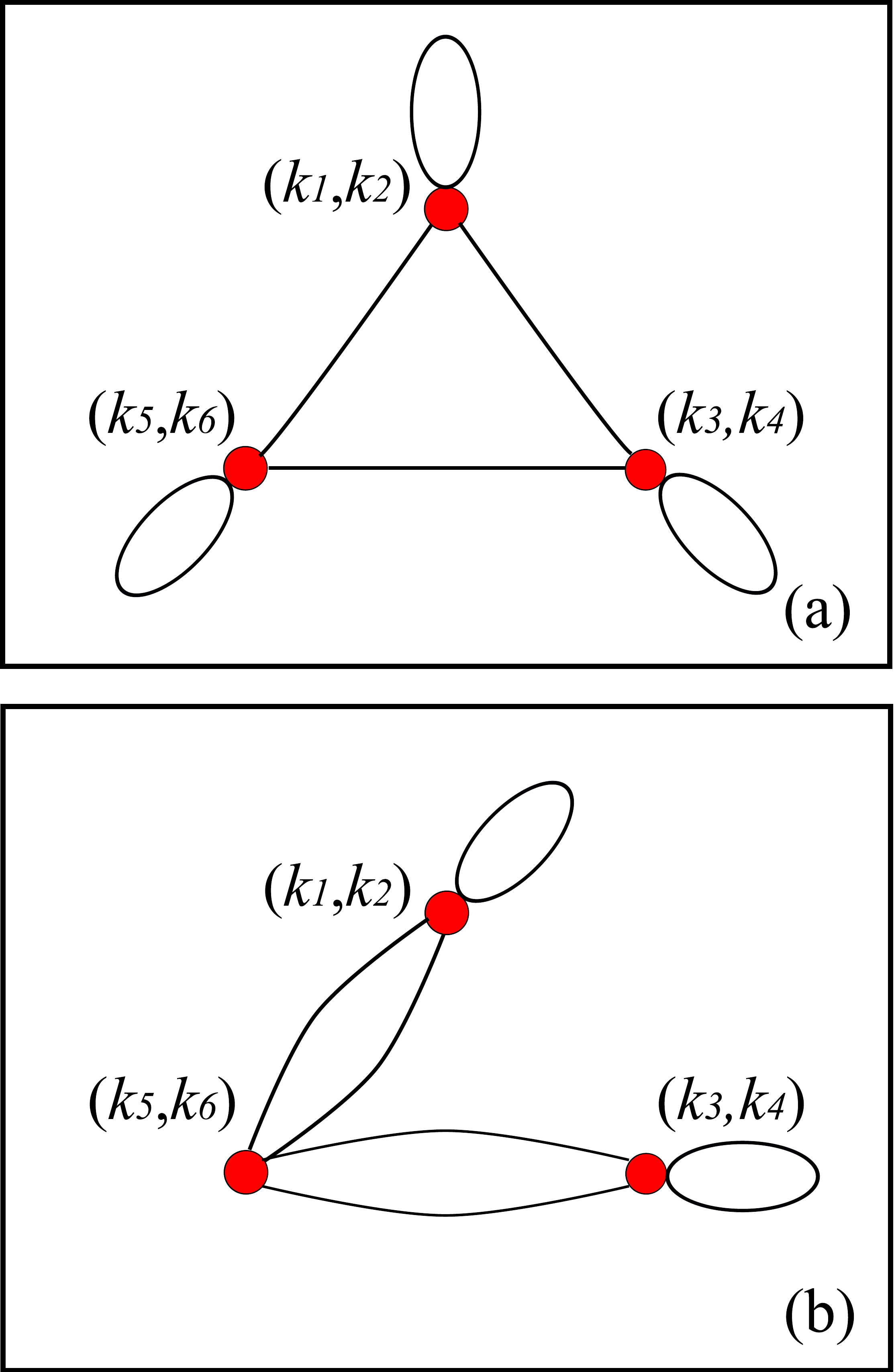}
\caption{(Color online) Connected graphs (with leaves) contributing to the second cumulant. (a): four-leaf cactus diagram. (b): four-leaf cactus diagram. (c): one leaf}
\label{fig:graphthirdcumconnected}
\end{center}
\end{figure}

\begin{figure}[t]
\begin{center}
\includegraphics[width=0.48\textwidth]{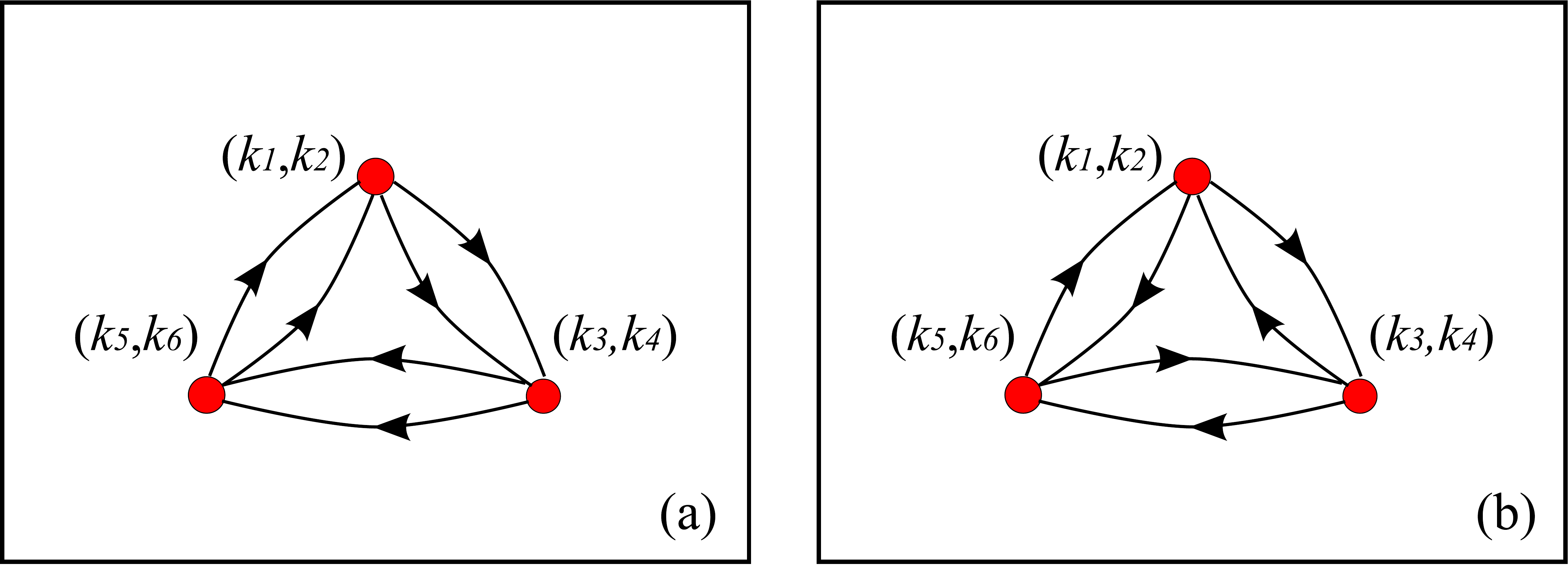}
\caption{(Color online) Connected graphs (without leaves) contributing to the second cumulant. 
(a): same internal and external orientation of the edges. (b): opposite internal and external orientation of the edges.}
\label{fig:graphthirdcumconnectednoloop}
\end{center}
\end{figure}

The contributions from connected graphs with three leaves (degeneracy $128$), represented in Fig.~\ref{fig:graphthirdcumconnected}(a), and with two leaves (degeneracy $192$), represented in Fig.~\ref{fig:graphthirdcumconnected}(b), are equal
\begin{equation}\label{eq:thirdcumgraphcactus}
[1\; 6, 3\; 2, 5\; 4] 
=[1\; 3, 2\; 5, 4\; 6] =N\frac{(N_A+N_{\bar{A}})^3}{8}\sim N^{5/2},  
\end{equation}
The graph represented in Fig.~\ref{fig:graphthirdcumconnected}(c) (degeneracy 192) gives an  asymptotic contribution
\begin{eqnarray}
\quad[1\; 6, 2\; 5, 3\; 4]\sim
3\sqrt{2}N^{\alpha+1/2} .
\end{eqnarray}
Finally, the asymptotic contributions from the graphs in Fig.\ref{fig:graphthirdcumconnectednoloop}(a) and (b) read, respectively \cite{classical}
\begin{eqnarray}
\quad[5\; 6, 1\; 2, 3\; 4]&\sim&N^{\alpha},\\
\quad[3\; 6, 5\; 2, 1\; 4]&\sim& c N^{\gamma},
\end{eqnarray}
with 
$c\simeq1.05385$ and $\gamma\simeq 1.8417$.
Notice the appearance of a new irrational exponent $\gamma$, which again, by cancellations, turns out to be the dominant contribution to the cumulant $\kappa_0^{(3)}$, followed by that due to $\alpha$. In summary, the asymptotic value is
\begin{equation}
\label{eq:k03A}
\kappa_0^{(3)}\simeq 67.4\ N^{-4.158...}.
\end{equation}

Again, by retaining only the contributions from Eq.~(\ref{eq:thirdcumgraphcactus}), as for $\tilde{\kappa}_0^{(2)}$, which dominate the large $N_c$ limit we find
\begin{eqnarray}\label{eq:third}
\tilde{\kappa}_0^{(3)} &=& \frac{40 (N_A+N_{\bar{A}})^3} {(N+1)(N+2)(N+3)(N+4)(N+5)}
\nonumber\\
&\sim&  \frac{320}{N^{7/2}}
\end{eqnarray}
where the asymptotic expression is obtained for balanced  bipartitions {of an even number of qubits},  $N_A=N_{\bar{A}}=\sqrt{N}$, and $N\rightarrow +\infty$.

For the fourth and fifth cumulant we have to resort to numerical methods. By generating a large number of random vectors for different $N$, we find that
\begin{eqnarray}
\label{eq:k0rN}
\kappa_0^{(3)}&=&(43\pm 12)N^{-4.18\pm0.06}\\
\kappa_0^{(4)}&=&(27\pm 20)N^{-5.2\pm0.5}\\
\kappa_0^{(5)}&=&(148\pm 90)N^{-6.5\pm1.5}.
\end{eqnarray}
where $\kappa_0^{(3)}$ is in good agreement with the theoretical value (compare (\ref{eq:k0rN}) with (\ref{eq:k03A})) .

These numbers suggest a scaling of the form $\kappa_0^{(n)}\sim N^{-0.25-1.4 n}$ (whose quality can be seen in figure), in partial agreement with the large-$N_c$ result is $\tilde\kappa_0^{(n)}\sim N^{1-1.5 n}$.

\begin{figure}[htbp]
\begin{center}
\includegraphics[width=0.9\columnwidth]{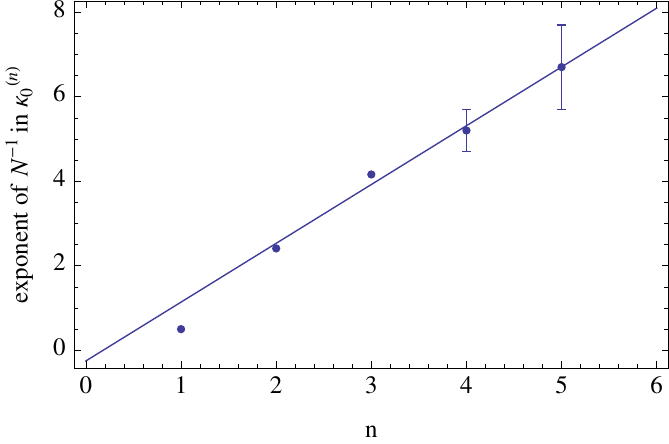}
\caption{The scaling exponent of $\kappa_n$ from numerics. On the first 3 points the error bars are smaller than the symbol size and the data agrees with the theory.}
\label{fig:expofkn}
\end{center}
\end{figure}

\section{Cactus diagrams and large number of colours}
\label{cactus}

As the reader should have deduced from the previous sections, the proliferation of diagrams and the variety of the {exponents} on $N$ appearing prevent one from writing a closed form for the correlation functions $\langle z_i \bar{z}_j\rangle_\beta$. One can then decide to sum a particular family of diagrams, as we did for the second and third cumulant in the high temperature expansion Eq.~(\ref{eq:second}) and Eq.~(\ref{eq:third}), hoping that this can give some information about the characterization of multipartite entanglement. 
However, the choice of the cactus (subdominant) diagrams was quite arbitrarily and motivated only by the fact that they yield integer exponents.

Of course, one should choose the family of graphs with an objective criterion, not on the basis of ease of computation. An objective criterion which is common practice in field theory is that of generalizing the model to a larger symmetry group, from $SO(2)$ to $SO(N_c)$, and then taking the limit of large number of colours $N_c$~\cite{qcd,parisi,jzj}. This will select a family of diagrams, and enable us to compute their amplitudes in a closed form to leading order in $1/N_c$. We will see that these diagrams are exactly the cactuses. Moreover, we will see, at the end of the computation, that a phase transition is obtained at a temperature below which a combination of the rotation group $SO(N_c)$ and the symmetric group $\cS_n$ is spontaneously broken.

In order to perform this calculation, we will rewrite the expression of the potential of multipartite entanglement in a different form.
The Fourier coefficients $z_k$ of a quantum state can be written as
\begin{equation}
z_k=\Phi_k^1+i\Phi_k^2,
\end{equation}
with $\Phi_k^{\mu}$, $\mu=1,2$, real numbers.
We will consider this object as a two components vector.
In terms of these real quantities, we can rewrite the Hamiltonian  in Eq.~(\ref{eq:pimeDelta}) 
 in the following form [using the symmetries in Eq.~(\ref{eq:symmetries})]
\begin{eqnarray}\label{eq:realhamiltonian}
H&=& \sum_{k,k',l,l' \in {\mathbb{Z}}_2^n}\Delta(k,k';l,l')
\nonumber\\&\times&
\left[2\sum_{\mu,\nu=1}^2\Phi_k^{\mu}\Phi_l^{\mu}\Phi_{k'}^{\nu}\Phi_{l'}^{\nu} - \sum_{\mu,\nu=1}^2\Phi_k^{\mu}\Phi_{k'}^{\mu}\Phi_{l}^{\nu}\Phi_{l'}^{\nu}\right] \nonumber\\
&=&\frac{N_c}{2} \sum_{k,k',l,l' \in {\mathbb{Z}}_2^n}\tilde{\Delta}(k,k';l,l')
\sum_{\mu,\nu=1}^{N_c}\Phi_k^{\mu}\Phi_l^{\mu}\Phi_{k'}^{\nu}\Phi_{l'}^{\nu} ,
\nonumber\\
\end{eqnarray}
for $N_c\equiv2$, and
\begin{equation}\label{eq:deltatilde}
\tilde{\Delta}(k,k';l,l')=2\Delta(k,k';l,l')-\Delta(k,l;k',l').
\end{equation}
This expression can be put in a more compact form by writing
\begin{equation}
\label{eq:realhamiltoniannc}
H
= \frac{N_c}{2} \sum_{k,k',l,l' \in {\mathbb{Z}}_2^n}\tilde{\Delta}(k,k';l,l')(\vec{\Phi}_k  \cdot \vec{\Phi}_l) (\vec{\Phi}_{k'}\cdot \vec{\Phi}_{l'}) , 
\end{equation}
where, in general, $\vec{\Phi}_{k}=(\Phi_k^1,\dots, \Phi_k^{N_c})$, and the dot denotes the scalar product. 
The index $\mu$, ranging form $1$ to $N_c$, will play in the following the role of a \emph{color} index, and we will be interested in the limit $N_c\to\infty$. The normalization of the complex vector $z_k$ becomes, for generic $N_c$ the constraint
\begin{equation}
\label{eq:unitnormNc}
\sum_{k \in {\mathbb{Z}}_2^n}\sum_{\mu=1}^{N_c} \left(\Phi_{k}^{\mu}\right)^2 =
\sum_{k \in {\mathbb{Z}}_2^n} \vec{\Phi}_{k}\cdot \vec{\Phi}_{k} = 1.
\end{equation}

We will now show that the Hamiltonian~(\ref{eq:realhamiltoniannc}) is positive for any color $N_c$, and  will find a lower bound which generalizes the bound on the potential of multipartite entanglement, valid for $N_c=2$.
 From~(\ref{eq:pimeDelta}) and the explicit expression of the purity across the bipartition $(A,\bar{A})$,
\begin{eqnarray}
\pi_A(z)= \sum_{k,k',l,l'} \delta_{k_{A}, l'_{A}} \delta_{k'_{A},l_{A}}
\delta_{k_{\bar{A}}, l_{\bar{A}}} \delta_{k'_{\bar{A}}, l'_{\bar{A}}} z_{k}\, z_{k'}\,
\bar{z}_{l}\, \bar{z}_{l'}\, , \quad 
\label{eq:Deltadef1}
\end{eqnarray}
it is not difficult to express the Hamiltonian as an average over bipartitions
\begin{equation}
H(\vec{\Phi}) = \left(\begin{array}{l}n
\\n_A\end{array}\!\!\right)^{-1}\sum_{|A|=n_A}H_{A}( \vec{\Phi} ) ,
\label{eq:Haverage}
\end{equation}
where
\begin{eqnarray}
H_A(\vec{\Phi}) &=& \frac{N_c}{2} \sum_{k,k'}\sum_{\mu,\nu}\left(2\Phi^{\mu}_{k_A k_{\bar{A}}} \Phi^{\mu}_{k'_A k_{\bar{A}}} \Phi^{\nu}_{k'_A k'_{\bar{A}}} \Phi^{\nu}_{k_A k'_{\bar{A}}} \right.
\nonumber\\
& & \qquad \qquad \quad \left.
- \Phi^{\mu}_{k_A k_{\bar{A}}} \Phi^{\nu}_{k'_A k_{\bar{A}}} \Phi^{\mu}_{k'_A k'_{\bar{A}}} \Phi^{\nu}_{k_A k'_{\bar{A}}}
\right), \nonumber\\
\label{eq:HAPhi}
\end{eqnarray}
with $k_A= (k_i)_{i\in A}$. 

Now, let
\begin{equation}
X_{k_{\bar{A}} \mu , l_{\bar{A}} \nu} = \sum_{k_A} \Phi^{\mu}_{k_A k_{\bar{A}}} \Phi^{\nu}_{k_A l_{\bar{A}}},
\label{eq:Xdef}
\end{equation}
and note that $X$ is a symmetric $N_{\bar{A}} N_c \times N_{\bar{A}} N_c$ matrix, namely $X^T = X$.
We get
\begin{equation}
H_A = \frac{N_c}{2} \sum_{k_{\bar{A}},l_{\bar{A}}}\sum_{\mu,\nu}\left(2 X_{k_{\bar{A}}\mu,l_{\bar{A}}\nu}^2  
- X_{k_{\bar{A}}\mu,l_{\bar{A}}\nu} X_{k_{\bar{A}}\nu,l_{\bar{A}}\mu}\right),
\end{equation}
that is
\begin{equation}
H_A = \frac{N_c}{2} \left[ 2\tr(X^{T} X) - \tr (X^{T} Y)\right],
\end{equation}
with 
$Y_{k_{\bar{A}} \mu , l_{\bar{A}} \nu}=X_{k_{\bar{A}} \nu , l_{\bar{A}} \mu}$ a symmetric matrix.
By the Cauchy-Schwarz inequality for the Hilbert-Schmidt scalar product we get
\begin{equation}
\tr (X^T Y) \leq \tr (X^T X)^{1/2} \tr (Y^T Y)^{1/2}.
\end{equation}
But it is easy to see that $\tr (Y^T Y) = \tr (X^T X)$, so that
\begin{equation}
H_A(\vec{\Phi}) \geq \frac{N_c}{2} \tr(X^2) >0.
\end{equation}
By making use  of the constraint~(\ref{eq:unitnormNc}) we can estimate the positive lower bound as follows.
First notice that the positive matrix $X\geq 0$ has unit trace. Indeed,
\begin{equation}
\tr X = \sum_{k_{\bar{A}},\mu} X_{k_{\bar{A}}\mu,k_{\bar{A}}\mu} =\sum_{k_A,k_{\bar{A}},\mu} \Phi^{\mu}_{k_A,k_{\bar{A}}} \Phi^{\mu}_{k_A,k_{\bar{A}}} =1,
\end{equation}
by~(\ref{eq:unitnormNc}). Therefore,
\begin{equation}
\tr (X^2) \geq \frac{1}{\mathrm{rank} X} \geq \frac{1}{N_c N_{\bar{A}}}.
\end{equation}
But, from the definition~(\ref{eq:Xdef}), we get that ${\mathrm{rank} X}\leq N_A$, so that
\begin{equation}
H_A(\vec{\Phi}) \geq \frac{N_c}{2 N_A}.
\label{eq:lowerbound}
\end{equation}
By a straightforward computation, one can check that the minimum is attained at
\begin{equation}
\Phi_{k}^\mu = \frac{\phi^\mu}{\sqrt{N_A}} \delta_{k_{\bar{A}}, k_A } , 
\qquad \sum_\mu (\phi^{\mu})^2 =1,
\label{eq:maxentPhi}
\end{equation}
where the Kronecker delta is meant to be $1$ when $k_{\bar{A}} = (k_A, 0,\dots,0)$. Notice that~(\ref{eq:maxentPhi}) is the direct generalization of  the maximally bipartite entangled state across the bipartition $(A,\bar{A})$, with coefficients $z_k = \e^{\i \alpha_k} \delta_{k_{\bar{A}}, k_A } /\sqrt{N_A}$. 

By plugging~(\ref{eq:lowerbound}) into~(\ref{eq:Haverage}) we finally get the desired lower bound
\begin{equation}
\min H \geq \frac{N_c}{2 N_A},
\end{equation}
which may not be attained due to frustration among the bipartitions. See the discussion in Sec.~\ref{sec:minimum}.

However, an interesting simplification occurs in the limit $N_c\to\infty$. 
We now introduce the constraint (\ref{eq:unitnormNc}) by means of a Lagrange multiplier $\lambda$ and rescale (for future purposes) the inverse fictitious temperature $\tilde{\beta}=\beta/\beta_0$, with
\begin{equation}
\label{eq:beta0}
\beta_0= \frac{ 2 N^2}{N_A + N_{\bar{A}}-1} \sim N^{3/2}.
\end{equation} 
We are finally left with the modified Hamiltonian
{\begin{eqnarray}
\tilde{\beta} \cH(\lambda)&=&
\tilde{\beta} \beta_0 \frac{N_c}{2}\sum_{k,k',l,l' }\tilde{\Delta}(k,k';l,l')(\vec{\Phi}_k  \cdot \vec{\Phi}_l) (\vec{\Phi}_{k'}\cdot \vec{\Phi}_{l'})
\nonumber\\
& & + \lambda \frac{N_c N}{2} \left(\sum_{k} \vec{\Phi}_k\cdot \vec{\Phi}_k-1\right).
\label{eq:hamiltNc}
\end{eqnarray}}

The partition function is an integral over $\Phi$ and over the Lagrange multiplier imposing the constraint so that
\begin{equation}
Z=\int \d\lambda\ \d\Phi\ e^{-\tilde{\beta} \cH(\lambda)}.
\end{equation}
The evaluation of $Z$ can be done by expanding it for small $\tilde{\beta}$ since the $\lambda$ part is quadratic and, after resummation of the diagrams, calculating the saddle point in $\lambda$. 

The saddle point equation in $\lambda$ is 
\begin{equation}
\frac{d}{d\lambda}\langle\cH\rangle_\beta=0
\end{equation}
and it is equivalent to the request that
\begin{equation}
\sum_{k}\langle \vec{\Phi}_k \cdot \vec{\Phi}_k \rangle_{\beta} =1,
\label{eq:avconst}
\end{equation}
where the average is evaluated using the full partition function $Z$.

We should find the value of $\lambda(\tilde{\beta})$ that satisfies the constraint. For example, for $\tilde{\beta}=0$ we have
\begin{equation}
G^{(0) \mu \nu }_{k l} = \langle \Phi_k^{\mu}\Phi_l^{\nu} \rangle_{0}=\frac{1}{\lambda N_c N} \delta^{\mu \nu}\delta_{k l},
\label{eq:freeprop}
\end{equation}
where $\mu,\nu$ are colour indices; in order to satisfy the constraint we need $\lambda=1$. 

\begin{figure}
\begin{center}
\includegraphics[width=0.95\columnwidth]{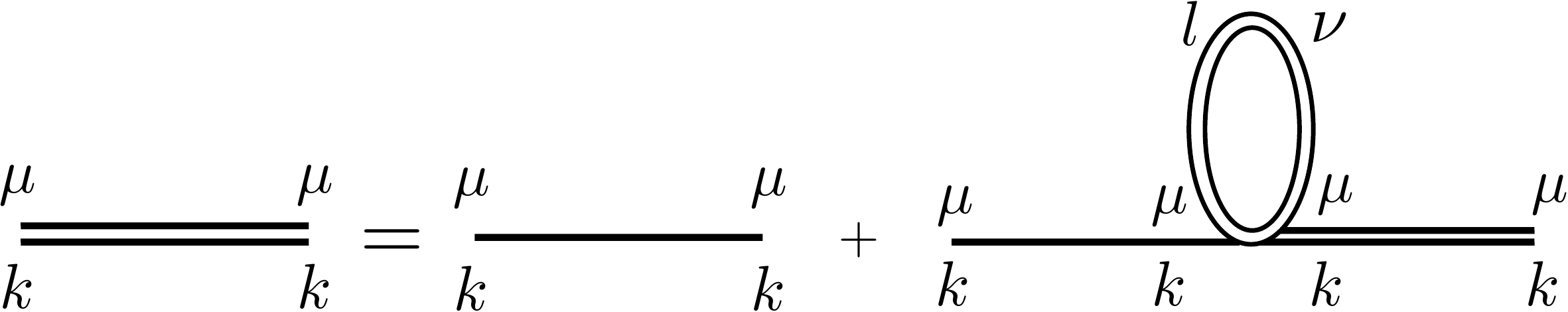}
\caption{(Color online) Dyson equation for the propagator in the large $N_c$ limit.}
\label{fig:Glargenc}
\end{center}
\end{figure}

In the limit $N_c\rightarrow\infty$ the diagrams giving the dominant contribution are the cactuses. The solution for $\tilde{\beta}>0$ can be obtained from the Dyson equation by considering only the non-zero propagator, obtained by setting $\mu=\nu,\ k=l$, as 
\begin{eqnarray}
G^{\mu\mu}_{kk} &=& \langle \Phi^{\mu}_k\Phi^{\mu}_k\rangle_{\beta}
 = G^{(0)\mu\mu}_{kk}
\nonumber\\
& & - \tilde{\beta}  \beta_0 \frac{N_c}{2} G^{(0)\mu\mu}_{kk}\sum_{\nu, l}\tilde{\Delta}(k,l;k,l)G^{\nu\nu}_{l l}G^{\mu\mu}_{kk} .
\nonumber\\
\label{eq:dyson}
\end{eqnarray}
From Eq.~(\ref{eq:freeprop}) we get
\begin{equation}\label{eq:gmumu}
G^{\mu\mu}_{kk} =\frac{1}{\lambda N_c N} -\frac{ \tilde{\beta} \beta_0 }{2 \lambda N}   \sum_{\nu, l}\tilde{\Delta}(k,l;k,l)G^{\nu\nu}_{l l}G^{\mu\mu}_{kk} .
\end{equation}
Let us define the quantity
\begin{equation}
G_l=\sum_{\nu=1}^{N_c}G^{\nu \nu }_{ll},
\end{equation}
that, inserted into Eq.~(\ref{eq:gmumu}), gives
\begin{equation}
G^{\mu\mu}_{kk} = \frac{1}{N_c N} \frac{1}{\lambda + \frac{\tilde{\beta} \beta_0}{2N}  \sum_l  \tilde{\Delta}(k,l;k,l) G_{l}}.
\end{equation}
We now sum over the colour index $\mu$ and obtain
\begin{equation}
G_{k} = \frac{1}{N} \frac{1}{\lambda +  \frac{\tilde{\beta} \beta_0}{2N}  \sum_l  \tilde{\Delta}(k,l;k,l) G_{l}}.
\label{eq:dyson2}
\end{equation}
We also notice that the constraint~(\ref{eq:avconst}) implies   
\begin{equation}
\sum_{k \in {\mathbb{Z}}_2^n} G_k=1.
\end{equation}

For sufficiently small $\tilde{\beta}$, we look for a solution with unbroken permutation symmetry ($\cS_N$-symmetric)
\begin{equation}
G_k=\frac{1}{N}, \qquad \forall k\in{\mathbb{Z}}_2^n,
\end{equation}
so that from Eq.~(\ref{eq:dyson2}) we have
\begin{equation}
\lambda + \frac{\tilde{\beta} \beta_0}{2N^2} \sum_l  \tilde{\Delta}(k,l;k,l) =1.
 \label{eq:lambdabeta}
\end{equation}
From Eq.~(\ref{eq:deltag}) and Eq.~(\ref{eq:deltatilde}) we get
\begin{eqnarray}
\sum_l  \tilde{\Delta}(k,l;k,l) &=&  \sum_{l} [2g(0,k\oplus l)-g(k\oplus l,k\oplus l)]
\nonumber\\
&=&  \sum_{l} [2g(0, l)-g(l,l)].
\end{eqnarray}
For balanced bipartitions 
we obtain
\begin{equation}
\sum_{l}g(0,l)=\frac{N_A+N_{\bar{A}}}{2},
\quad
\sum_{l}g(l,l)=1,
\end{equation}
so that
\begin{equation}
\sum_l  \tilde{\Delta}(k,l;k,l) =  N_A+N_{\bar{A}} -1
\label{eq:sumkdelta}
\end{equation}
independent of $k$, and Eq.~(\ref{eq:lambdabeta}) reads
\begin{equation}
\lambda  =1 -\tilde{\beta}.
\label{eq:lambdabeta1}
\end{equation}

\comm{In order to find the free energy we need to correct $f_0=-\ln Z_0$ with all the connected diagrams generated by $\langle e^{-\tilde{\beta} H_1}\rangle_0$. Again, working at leading order in $1/N_c$, we select the cactus diagrams and  find 
\begin{equation}
\tilde{\beta}\delta F= -\ln \langle e^{-\tilde{\beta} H_1}\rangle_0= \frac{\tilde{\beta}\beta_0 N_c}{2}\sum_{\mu,\nu,k,l}\tilde{\Delta}(k,l;k,l)G^{\nu\nu}_{l l}G^{\mu\mu}_{kk}.
\end{equation}

By inserting the previously found value of $G_{kk}^{\mu\mu}=1/N_cN$, we find
\begin{equation}
\tilde{\beta}\delta F
=\tilde{\beta} N_c N.
\end{equation}
Summing this result with $f_0$ evaluated in Eq.~(\ref{eq:freeen}), we find the value of the free energy as a function of $\tilde{\beta}$
\begin{eqnarray}
\tilde{\beta} F&=&-\ln Z=-\ln Z_0|_{\lambda=1-\tilde{\beta}}+\tilde{\beta} N_cN\nonumber\\
&=&-\frac{N_cN}{2}\left((1-\tilde{\beta})-\ln(1-\tilde{\beta})-2\tilde{\beta}\right).
\end{eqnarray}
Moreover, the expectation value of the potential of multipartite entanglement reads
\begin{equation}
\langle H_1\rangle_{\beta}=\frac{\partial (\tilde{\beta} F)}{\partial \tilde{\beta}}=N_c N \frac{2-3\tilde{\beta}}{2-2\tilde{\beta}}.
\end{equation}
Going back to $\beta=\tilde{\beta}\beta_0$ and $H=H_1/\beta_0$ we find
\begin{equation}
\langle H\rangle_\beta=\frac{N_c(2\sqrt{N}-1)}{2N}\frac{2-3\beta/\beta_0}{2-2\beta/\beta_0}.
\end{equation}
}

By using this result we get the average purity (again retaining only the leading order in $1/N_c$)
\begin{eqnarray}
\langle H\rangle_{\beta}&=&\frac{N_c}{2}\sum_{k,l,\mu,\nu}\tilde\Delta(k,l,k,l)G^{\mu\mu}_{kk}G^{\nu\nu}_{ll}
\nonumber\\
&=&\frac{N_c(N_A+N_{\bar{A}} -1)}{2N}
\sim \frac{N_c}{\sqrt{N}},
\label{eq:enbetacostant1}
\end{eqnarray}
independent of $\tilde\beta$ which, for $N_c=2$, gives the correct result for the value of the first cumulant in Eq.~(\ref{eq:firstcumul1}) only at $\tilde\beta=0$. This is due to the fact that, to the lowest order in $\tilde\beta$, sub-leading diagrams in $N_c$ are sub-leading in $N$ as well. Notice how, in this approximation, the dependence on the temperature has disappeared (a similar phenomenon occurs in a 
matrix model related to spin glasses \cite{CKPR}).

However, $\tilde\beta=1$ is a critical temperature, as one can see that the $\Phi$ fluctuations become massless. In fact, the {Lagrange} multiplier $\lambda$ is {the} coefficient of the quadratic part of the hamiltonian (\ref{eq:hamiltNc})
\begin{equation}
\frac{\partial^2 \cH}{\partial \Phi_k^\alpha\partial\Phi_l^\beta}\Bigg|_{\Phi=0}=\frac{1}{\tilde\beta}N_cN\delta^{\alpha\beta}\delta_{kl}\lambda.
\end{equation}
Therefore for $\tilde\beta>1$, $\lambda<0$ and for $N\to\infty$ we should expect spontaneous symmetry breaking such that for some $k$ $\langle\Phi_k\rangle>0$. The $\cS_N$ symmetry gets spontaneously broken.

This value of $\tilde\beta_c$ is not in evident agreement with the numerics, although a tendency {for} large $N_c$ of developing a kink at $\tilde\beta\sim 1$ is noted in the data for $n={3,} 4,5$ and $7$. In particular in Figs.~\ref{fig:3qubit}, \ref{fig:4qubit}, \ref{fig:5qubit} and \ref{fig:7qubit} one notices that for small $\tilde\beta$ by increasing $N_c$ the data move towards the large-$N_c$, small $\tilde{\beta}$, $\tilde{\beta}$-independent result (\ref{eq:enbetacostant1}), while for large $\tilde\beta$ the flow is reversed, probably asymptoting to the large-$N_c$ minimum value of the {Hamiltonian.}

A complete solution of the partition function of the quartic Hamiltonian (\ref{eq:hamiltNc}) and the $1/N_c$ corrections will be the subject of future work.


\section{Minimum of the Hamiltonian in the large-$N_c$ limit}
\label{sec:minimum}

In this section we will give an explicit example of what happens to the potential of multipartite entanglement for increasing values of $N_c$. 

Let us start by considering a collection of qubits. We recall that this case corresponds to $N_c=2$. In particular, it is well known \cite{scott,mmes,frustration}
that for $n=4$ or $n\ge 8$, the ideal minimum of the Hamiltonian (\ref{eq:realhamiltoniannc}) cannot be reached i.e.
\begin{equation}
E_0^{(n)}=\mbox{min} H \geq\frac{N_c}{2 N_A}=\frac{N_c}{2} 2^{-\left[\frac{n}{2}\right]},
\end{equation}
where $[.]$ denotes the integer part. We encounter the so-called \emph{frustration of multipartite entanglement}. In other words the requirement that the bipartite entanglement is maximal (minimal purity) for all bipartitions can engender conflicts. {Incidentally, we notice that this phenomenon can be found not only for spin systems  but also for infinite-dimensional systems as in the case of Gaussian states \cite{Gaussian1}.}

\begin{figure}[h]
\includegraphics[width=0.9\columnwidth]{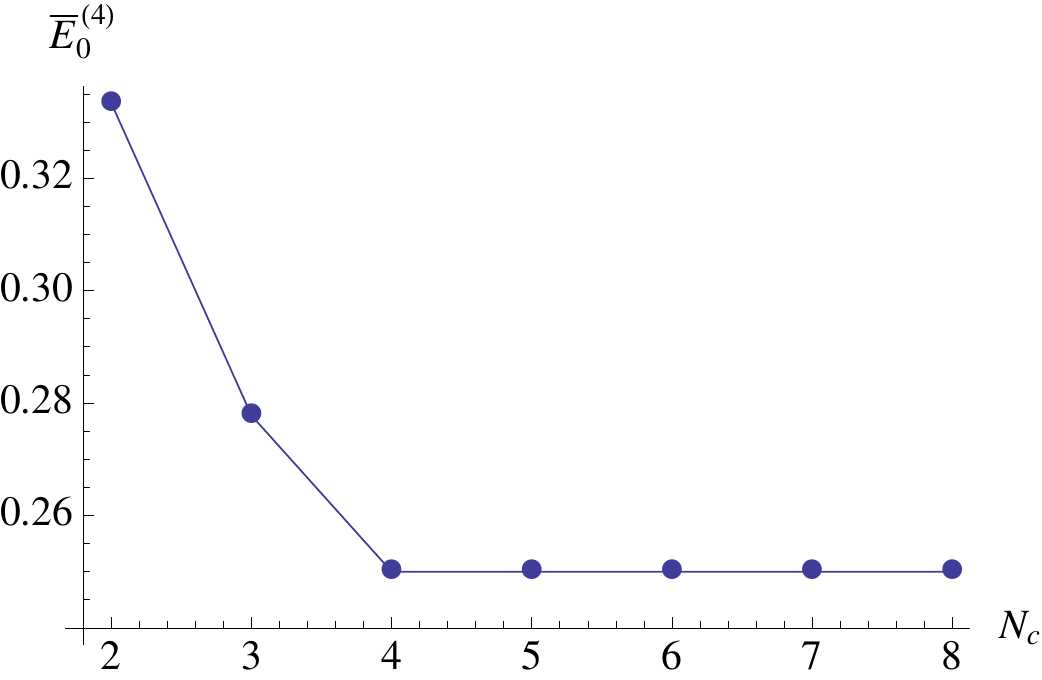}
\caption{Dependence of the rescaled minimum $\bar{E}_0^{(n)}$ of the Hamiltonian (\ref{eq:realhamiltoniannc}) on the value of $N_c$ for $n=4$.}
\label{fig:frust}
\end{figure}

In order to study this phenomenon, let us define the quantity
\begin{equation}
\bar{E}_0^{(n)}=\frac{2}{N_c} E_0^{(n)},
\end{equation}
which represents the rescaled minimum of the Hamiltonian (\ref{eq:realhamiltoniannc}) for different values of $N_c$. In order to understand the consequence of the large $N_c$ limit, we have performed a numerical minimization of Eq.~(\ref{eq:realhamiltoniannc}) for $n=4$ and $2\le N_c \le 8$.  In Fig.~\ref{fig:frust} we plot $\bar{E}_0^{(n)}$ as a function of $N_c$. It is manifest that by increasing the value of the colour parameter frustration disappears. Indeed, for $N_c=2$ we find $\bar{E}_0^{(n)}=1/3$, in agreement with previous results~\cite{gourwallach,HS00,BSSB}. If $N_c\ge4$ we have $\bar{E}_0^{(n)}=1/4$. Apparently, the ideal minimum is obtained by minimizing each term~(\ref{eq:HAPhi})
 in the Hamiltonian separately as if they were independent. Therefore for $N_c\geq 4$ the system is \emph{unfrustrated}. On the other hand, this means that one of the most characteristic trait of multipartite entanglement cannot be analyzed in the large-$N_c$ limit.

Incidentally, it is interesting to notice that for $n=4$ the values of the minimum energy follow the law
\begin{equation}
\bar{E}_0^{(4)}=\frac{N_c+2}{6 N_c}
\end{equation}
and that for $N_c\geq 4$ this expression becomes $\leq 1/4$.


\section{Search for the phase transition}
\label{sec:numeric}

\begin{figure}
\begin{center}
\includegraphics[width=\columnwidth]{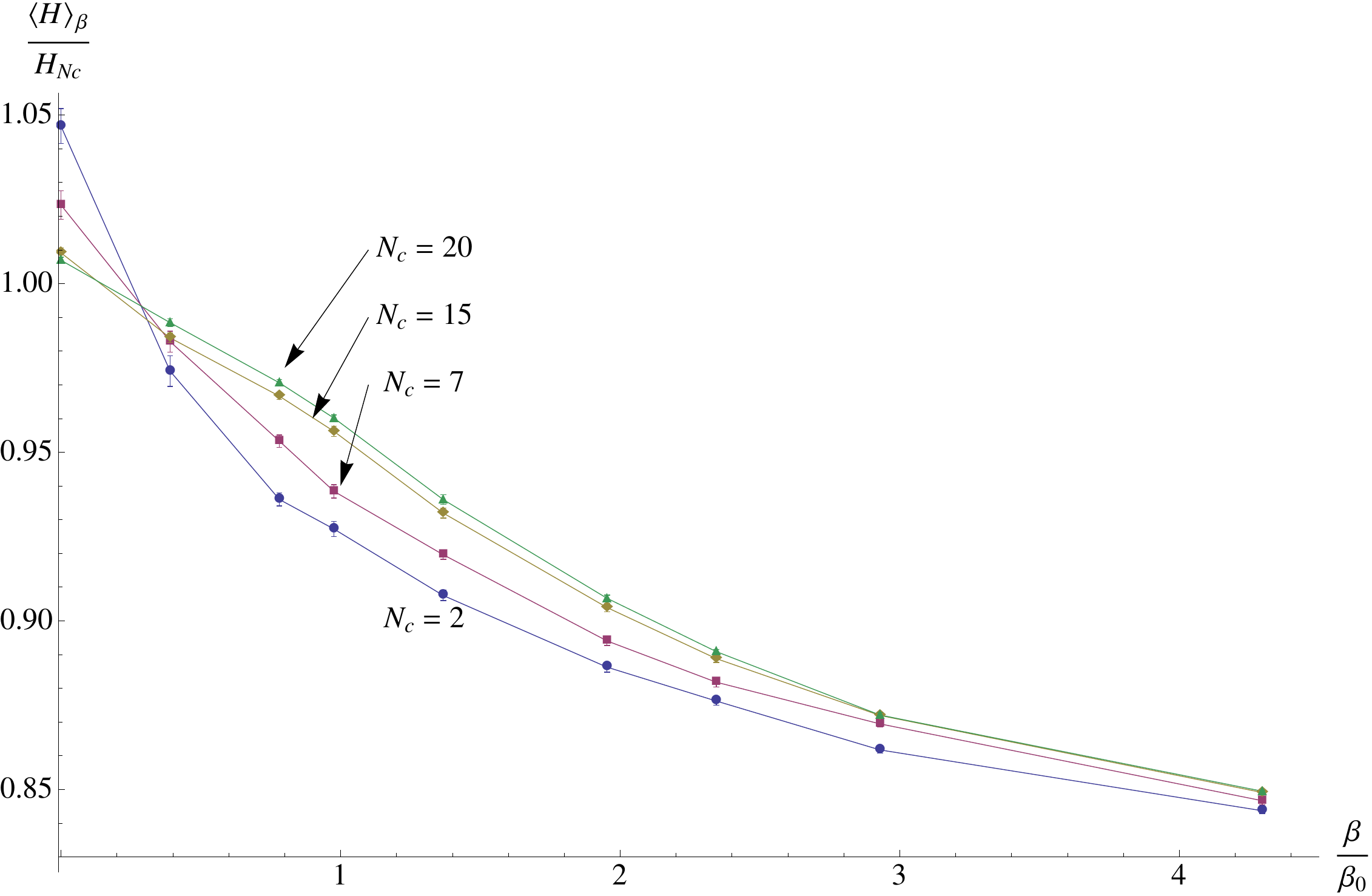}
\caption{(Color online) Numerical results for $n=3$ {($N=8$)} and $N_c$ ranging from 2 to 20.  $\beta_0$ and $H_{Nc}$ are defined {in} Eq.~(\ref{eq:beta0}) and Eq.~(\ref{eq:rescalH}).}
\label{fig:3qubit}
\end{center}
\end{figure}

\begin{figure}
\begin{center}
\includegraphics[width=\columnwidth]{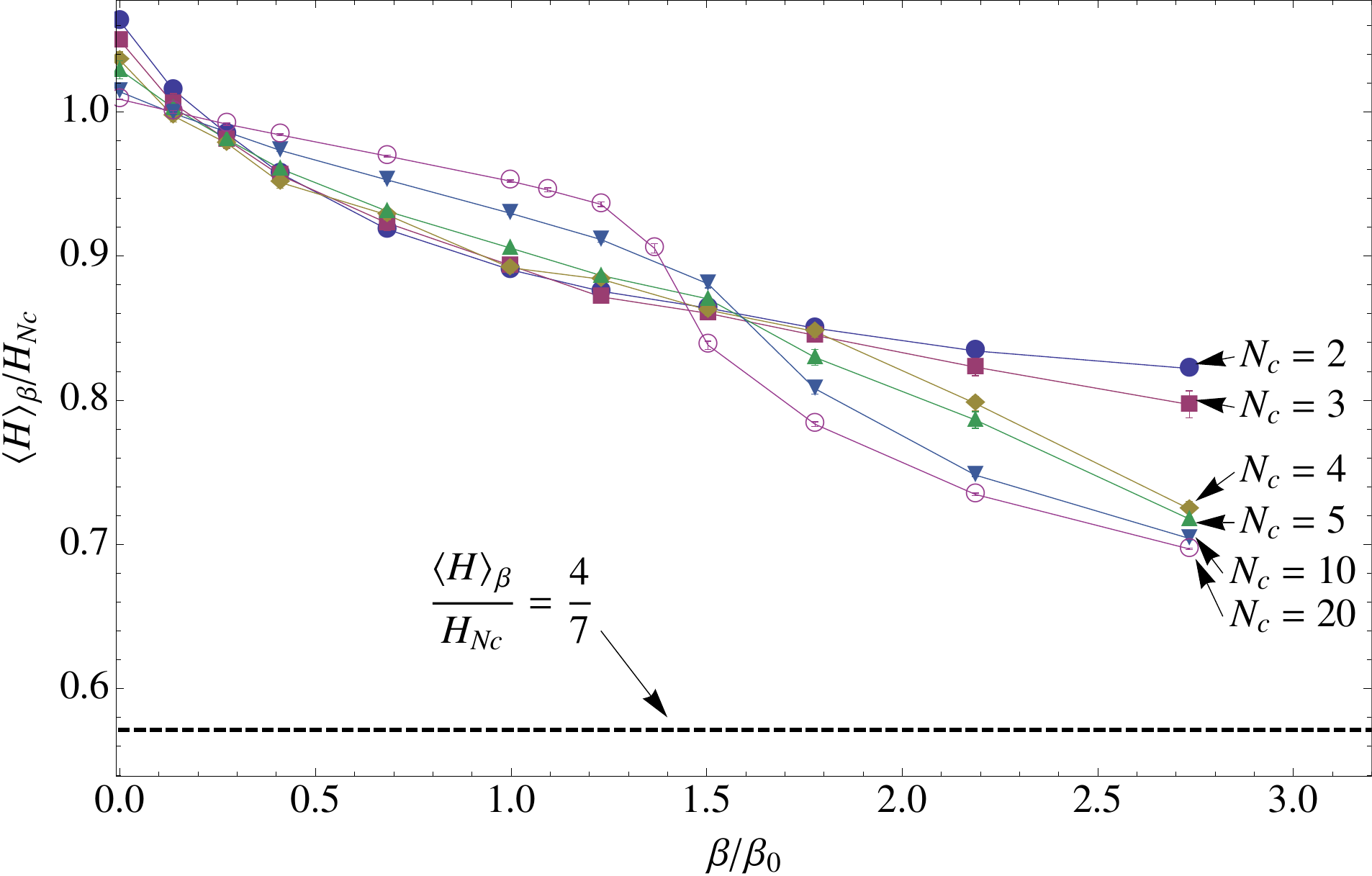}
\caption{(Color online) Numerical results for $n=4$ {($N=16$)} and $N_c$ ranging from 2 to 20.  $\beta_0$ and $H_{Nc}$ are defined {in} Eq.~(\ref{eq:beta0}) and Eq.~(\ref{eq:rescalH}).}
\label{fig:4qubit}
\end{center}
\end{figure}

\begin{figure}
\begin{center}
\includegraphics[width=\columnwidth]{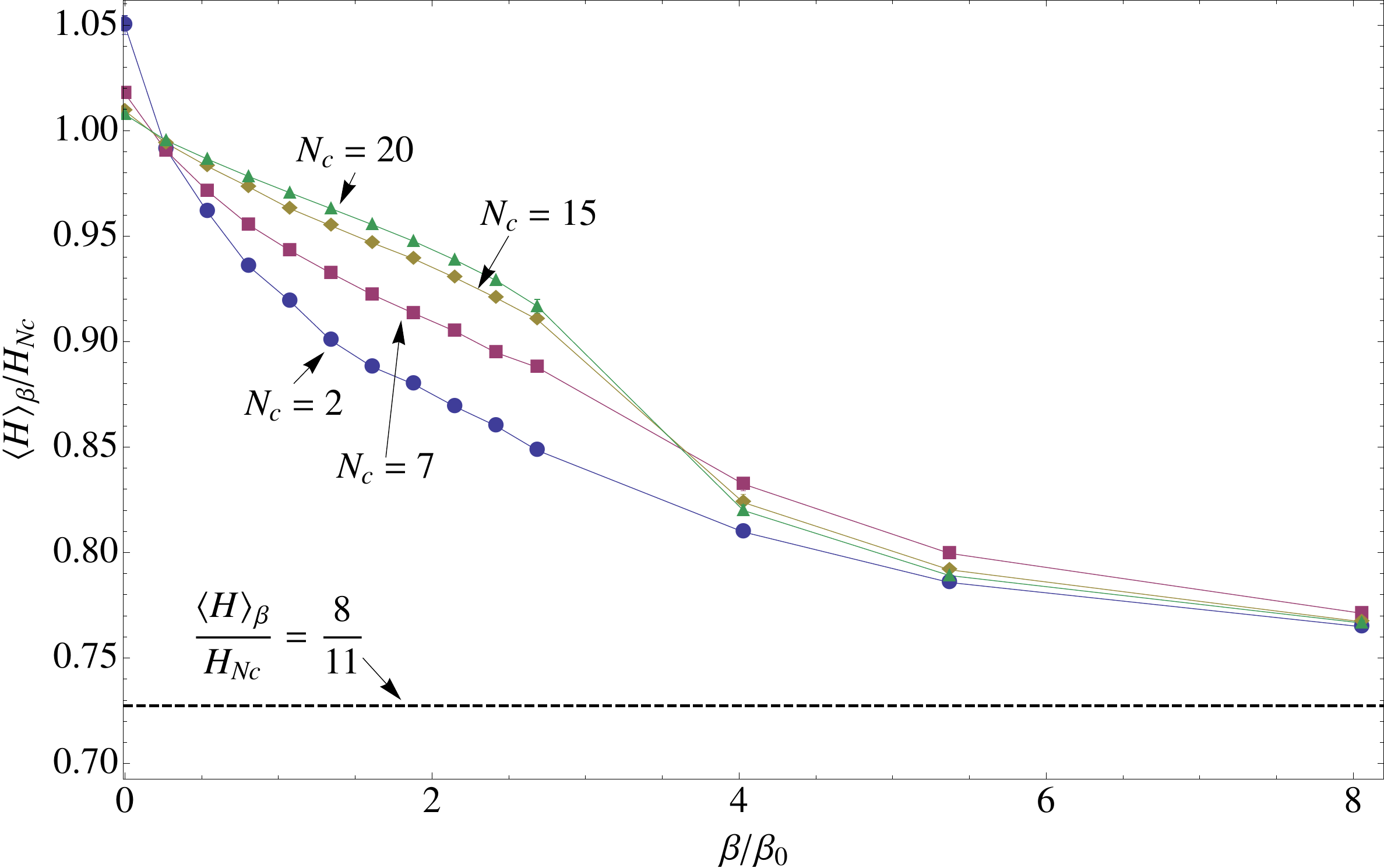}
\caption{(Color online) Numerical results for $n=5$ {($N=32$)} and $N_c$ ranging from 2 to 20. $\beta_0$ and $H_{Nc}$ are defined {in}  Eq.~(\ref{eq:beta0}) and Eq.~(\ref{eq:rescalH}). }
\label{fig:5qubit}
\end{center}
\end{figure}

\begin{figure}
\begin{center}
\includegraphics[width=\columnwidth]{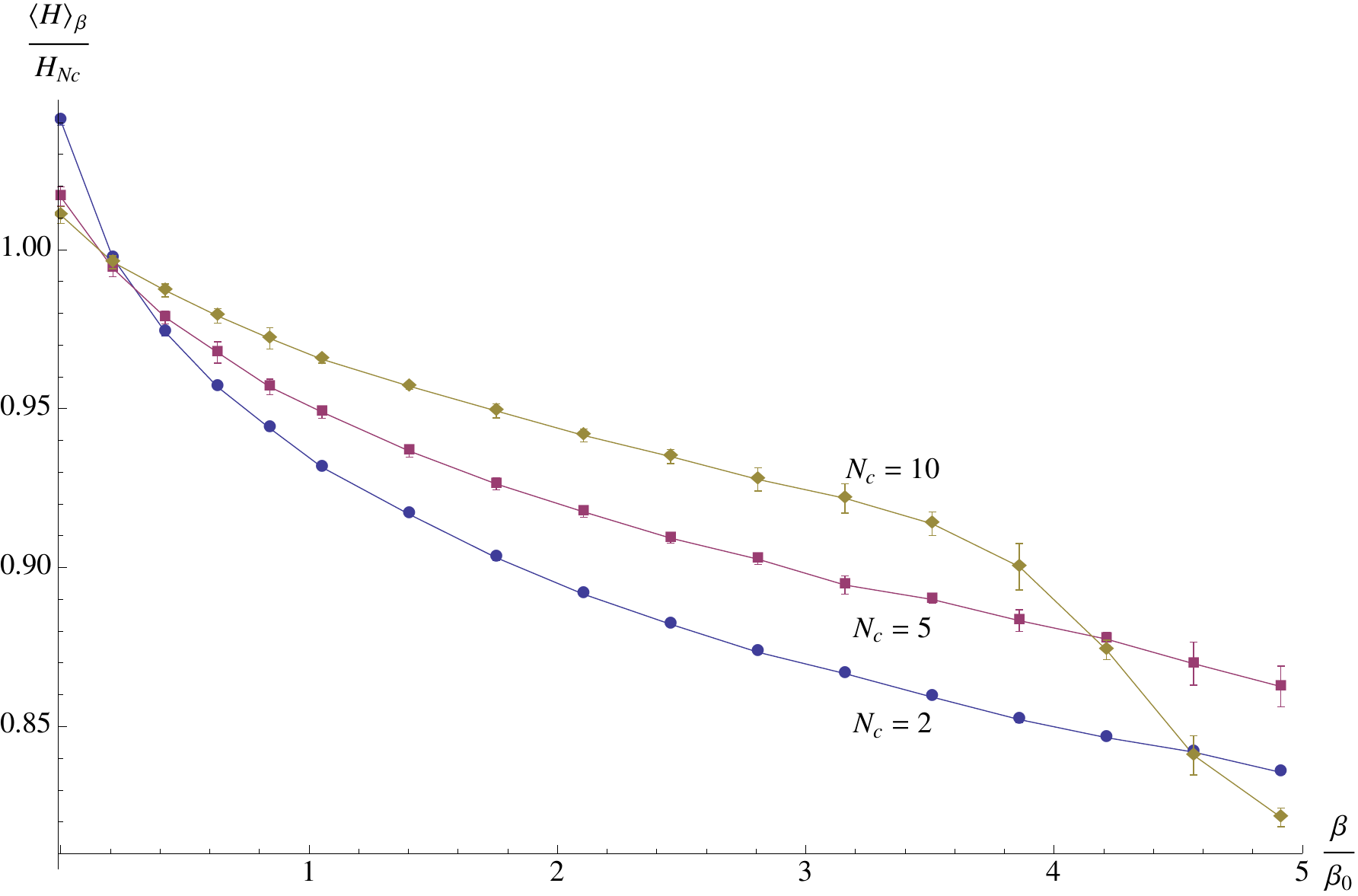}
\caption{(Color online) Numerical results for $n=7$ {($N=128$)} and $N_c$ ranging from 2 to 10. $\beta_0$ and $H_{Nc}$ are defined {in} Eq.~(\ref{eq:beta0}) and Eq.~(\ref{eq:rescalH}). }
\label{fig:7qubit}
\end{center}
\end{figure}

In Figures~\ref{fig:3qubit}, \ref{fig:4qubit}, \ref{fig:5qubit} and \ref{fig:7qubit} [respectively for $n=3$ {($N=8$)}, $n=4$ {($N=16$)}, $n=5$ {($N=32$)}, and $n=7$ {($N=128$)}] we plot the expectation value $\langle H\rangle_{\beta}$ rescaled by 
\begin{equation}
H_{Nc}=\frac{N_c(N_A+N_{\bar{A}} -1)}{2N}\label{eq:rescalH}
\end{equation}
versus $\tilde{\beta}=\beta/\beta_0$ for different values of the color number $N_c$ obtained with MonteCarlo simulations. We recall that $H$ is defined in Eq.~(\ref{eq:realhamiltoniannc}) and $\beta_0$ in Eq.~(\ref{eq:beta0}).

In all cases we notice that for large values of $N_c$ there is an inflection point. Moreover, for increasing values of $N_c$, the curves tend to become flatter for small values of  $\beta/\beta_0$. This behavior is more evident for $n=4$ {($N=16$)} and $n=7$ {($N=128$)}. Finally, we notice that for $n=4$ with $N_c > 3$  and $n=7$ with $N_c =10$ there is an evident change in the behavior of the curves for larger values of $\beta/\beta_0$, apparently absent in the case of $n=5$. This could be due to the removal of the {multipartite entanglement frustration} in the case $n=4$ as  {detailed in the previous} section (the frustration phenomenon is not present for $n=5$, and the case $n=7$ is still open). 

The flattening in the region of small $\tilde{\beta}=\beta/\beta_0$ might be interpreted as a confirmation of the presence of a phase transition around $\tilde{\beta}=1$ in the limit of large $N_c$. For $n=4$ (where this effect is more evident) we notice that the inflection point of the curves moves from values of $\tilde{\beta}\simeq 2$ to $\tilde{\beta}\simeq 1.5$.


\section{Search for hysteresis}
\label{sec:hysteresis}

We now try and elucidate the features of the phase transition brought to light in the preceding section. {In particular, we will search for the presence of hysteresis when the system is ``cooled'' fixing a large value of $\beta$ and, after equilibration, is taken back to the initial value of $\beta$. In order to perform this analysis we will} use  simulated-annealing-like algorithm~\cite{anneal1,anneal2}. {After fixing} an initial value of $\beta$ {we} let the system reach the equilibrium. At this point we start to ``heat'' the system and decrease the value of $\beta$ fixing the number of MonteCarlo steps before the following decrease.

\begin{figure}
\begin{center}
\includegraphics[width=\columnwidth]{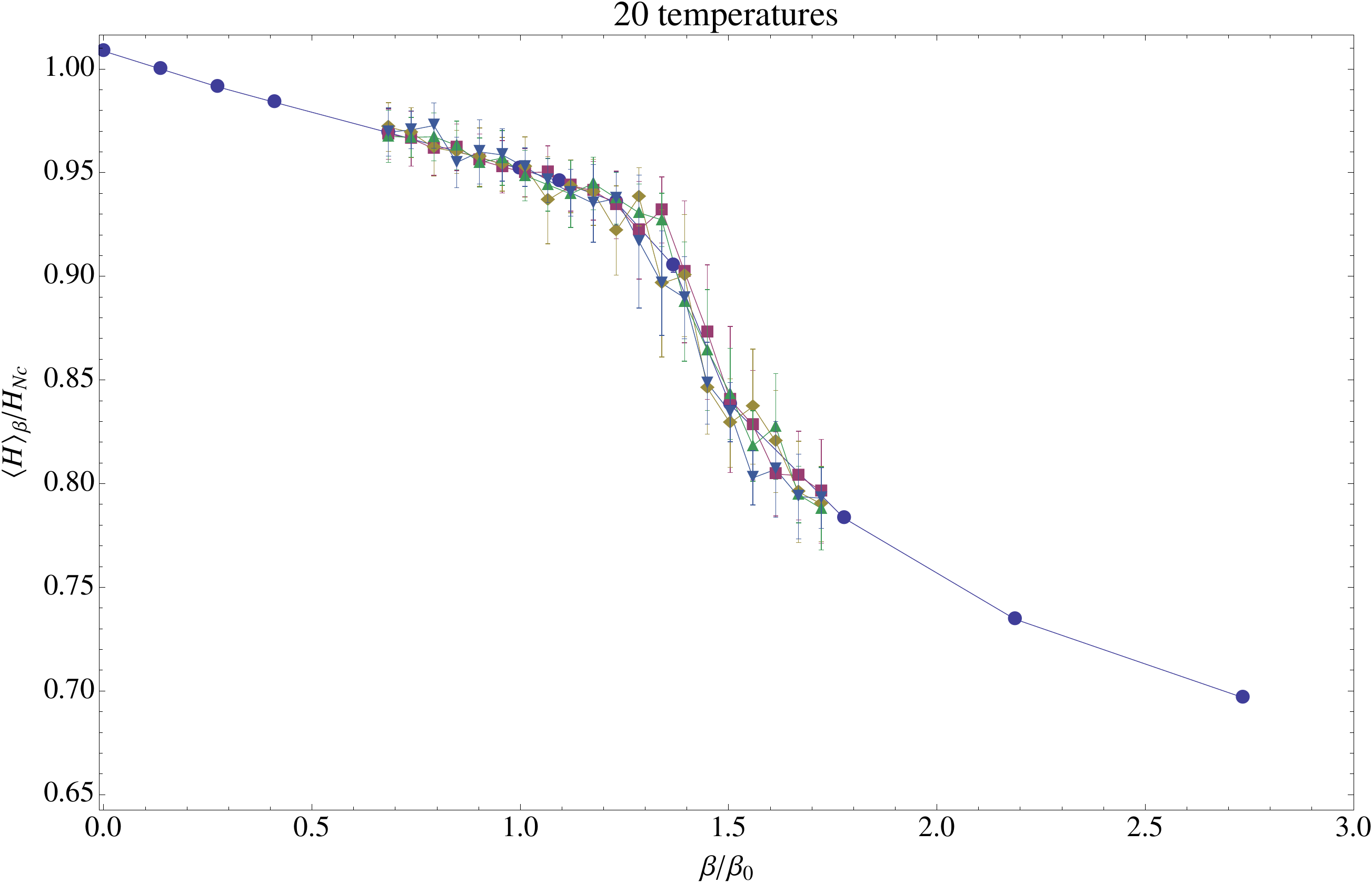}
\caption{(Color online) Search for the hysteresis phenomenon. Numerical results for $n=4$ {($N=16$)}, $N_c=20$. $\beta_0$ and $H_{Nc}$ are defined {in} Eq.~(\ref{eq:beta0}) and Eq.~(\ref{eq:rescalH}). Different curves correspond to different number of MonteCarlo steps between the decrease of $\beta$ in the simulated annealing procedure. See the text for details.}
\label{fig:temp20}
\end{center}
\end{figure}

In Fig.\ \ref{fig:temp20} we show the results obtained for the case $n=4$, $N_c=20$. We start from $\beta=130$ (point not shown in the figure), corresponding to $\beta/\beta_0\simeq 1.777$, and anneal for 20 values of $\beta$ (each step being equal to {$\Delta \beta=4$}). The different curves correspond to 50, 100, 200 and 300 Monte Carlo steps before each decrease of $\beta$. Notice that the second procedure is what one calls a quench in the literature (at variance with  annealing).
{We cannot see any} evidence of hysteresis. {As a reference, we have included in Fig.\ \ref{fig:temp20} also the curve (blue dots) that we have already included in Fig.\ \ref{fig:4qubit}: we remember that each point has been obtained not with simulated annealing but starting from $\beta=0$, fixing the value of $\beta$, let the system reach the equilibration and then performing a Monte Carlo run;} the other curves in Fig.\ \ref{fig:temp20} coincide within errors. The annealing was repeated for 10 and 5 values of $\beta$, with no difference and no evidence of hysteresis.

\begin{figure}
\begin{center}
\includegraphics[width=\columnwidth]{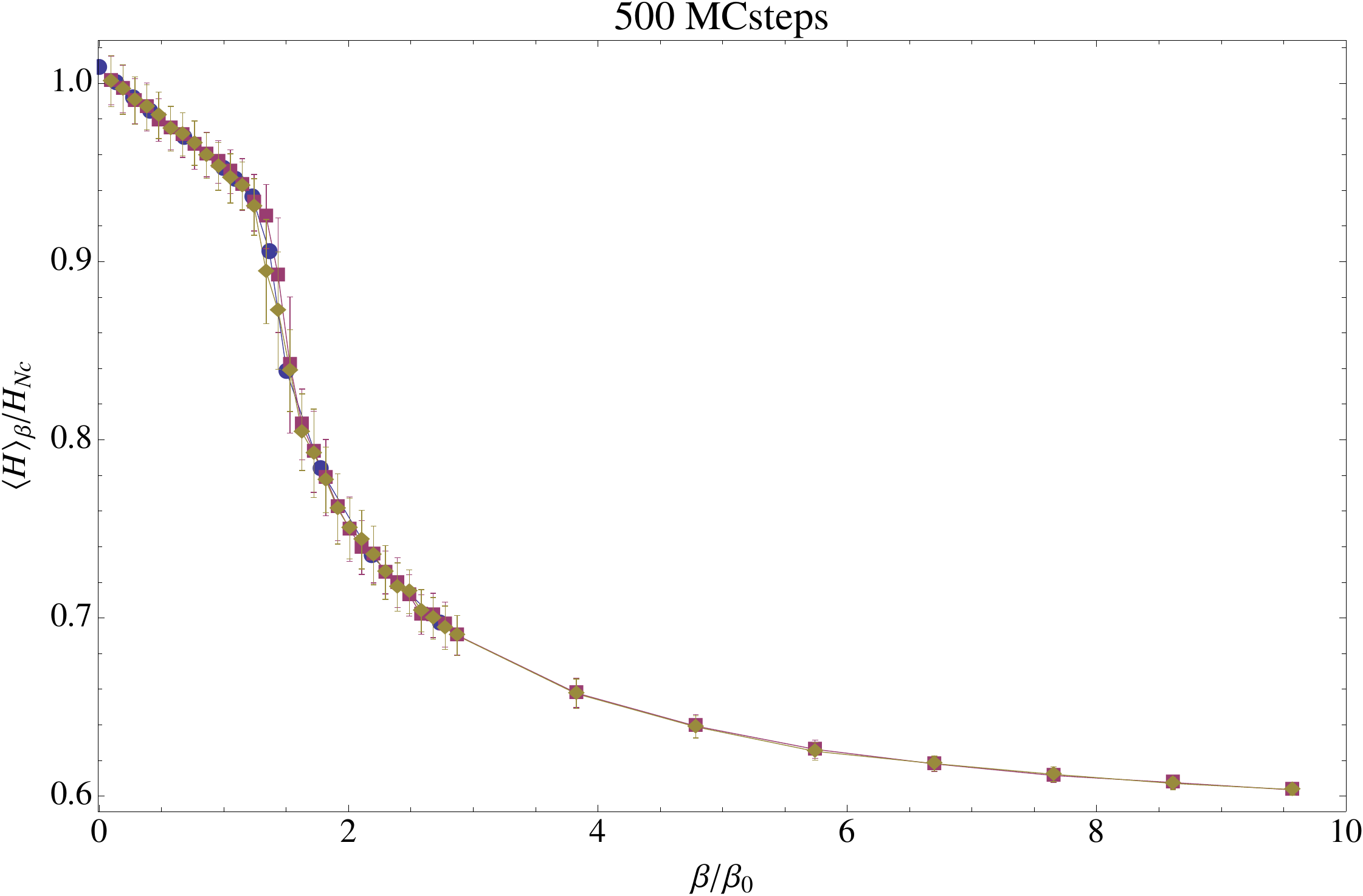}
\caption{(Color online) Numerical results for $n=4$ {($N=16$)}, $N_c=20$, and $500$ MonteCarlo steps. $\beta_0$ and $H_{Nc}$ are defined {in} Eq.~(\ref{eq:beta0}) and Eq.~(\ref{eq:rescalH}). See the text for details.}
\label{fig:mc500}
\end{center}
\end{figure}

In order to compare the results using different procedures, we have tried to perform a simulated annealing procedure also going from $\beta=0$ to larger values of the inverse temperature. We have changed the speed of the annealing procedure by varying the number of Monte Carlo steps between successive temperatures. The system is cooled starting from a given temperature, without going back to the initial state at every step. We expect to find the same results as in Fig.\ \ref{fig:temp20}.
We start from $\beta/\beta_0 =0$ and then proceed with simulated annealing at steps $\simeq 0.1$ up to {$\beta/\beta_0 = 3$}
and then at steps $\simeq 1$ up to {$\beta/\beta_0 = 10$.}
The rescaled energy for 500 Monte Carlo steps between different temperatures is plotted in Fig.\ \ref{fig:mc500}, {where we show (blue dots) the curve obtained without the simulated annealing and two different simulations with the simulated annealing (yellow and purple squares).} The results corroborate our previous finding and extend the graph to higher values of $\beta$ (below the frustrated minimum).
No difference is observed for 100 Monte Carlo steps (not shown). We conclude that no hysteresis is present.

\begin{figure}
\begin{center}
\includegraphics[width=\columnwidth]{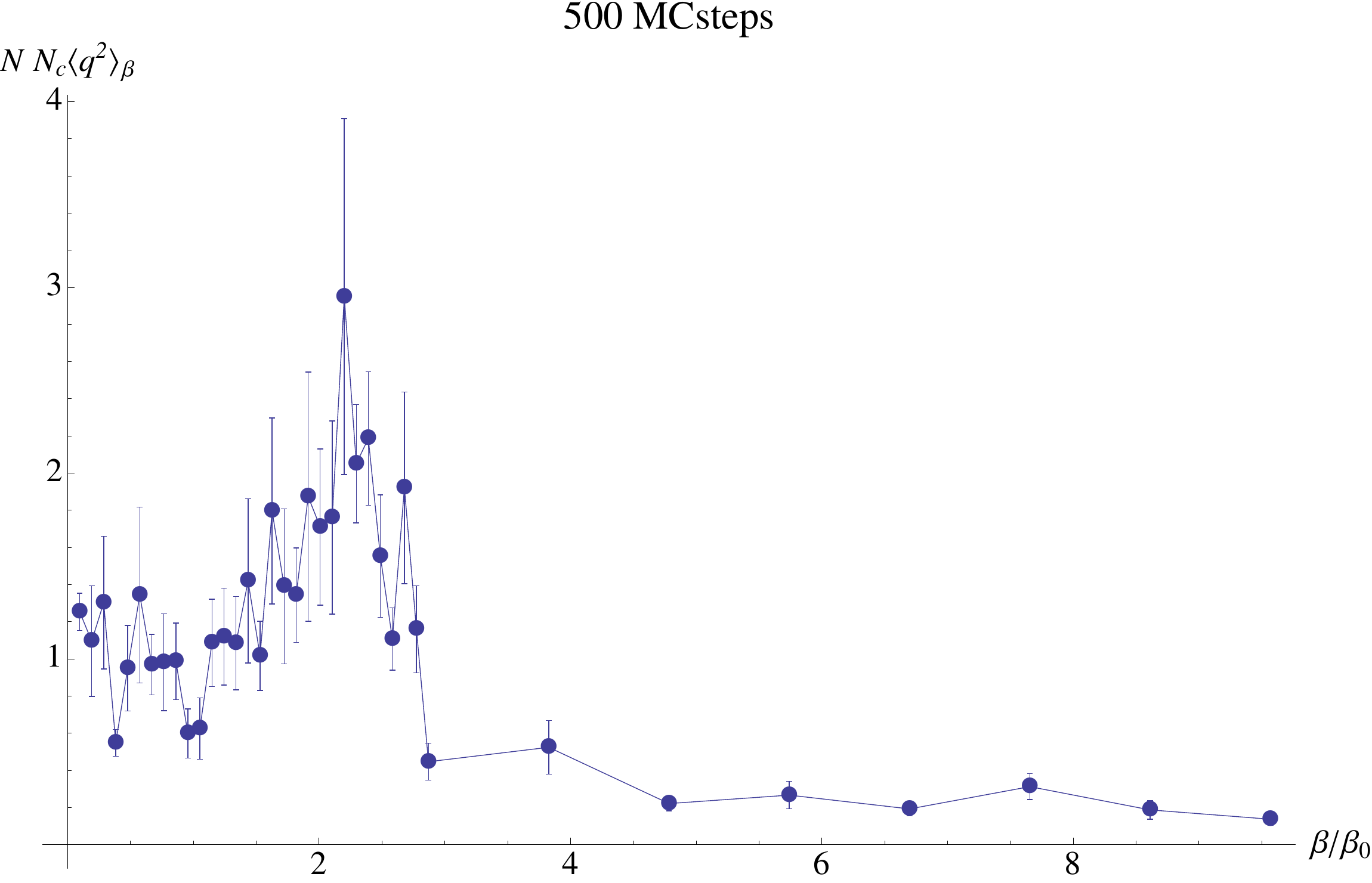}
\caption{(Color online) Overlap of configurations obtained from two different Monte Carlo simulations. The number of steps between different successive values of $\beta$ is 500. See the text for details.}
\label{fig:overlap500}
\end{center}
\end{figure}

As a final test, we have considered the behavior of the \emph{overlap} between configurations of two different Monte Carlo simulation run in parallel (labelled 1 and 2, respectively). This quantity is defined as
\begin{equation}
{\langle q^2\rangle}_\beta=\bigg\langle \bigg[\sum_{k} 
\vec{\Phi}_{k}^{(1)} \cdot \vec{\Phi}_{k}^{(2)}
\bigg]^2\bigg\rangle_\beta ,
\label{eq:overlap}
\end{equation} 
where the average is performed at a fixed value of $\beta$, by considering a configuration every ten Monte Carlo steps. We have considered the case of  500 Monte Carlo steps between different values of $\beta$. The results are shown in Fig.~\ref{fig:overlap500}, where each point is obtained averaging over 50 configurations at fixed $\beta$. The vertical axis is rescaled by the quantity $1/N N_c$ which corresponds to the value of two random independent configurations.
There is a decrease of the overlap around $\beta/\beta_0\simeq 2.5$, which most likely corresponds to different replicas freezing in different minima. However, these results cannot be considered decisive for the presence of a phase transition.

\section{Conclusions}
\label{sec:conclusions}

We have investigated the behavior of the potential of multipartite entanglement by generalizing the problem to $N_c$ colors and taking a large-$N_c$ limit. In the analytical treatment we find that an instability occurs for sufficiently small temperatures that breaks the permutational symmetry between spin configurations. This instability generates, in the thermodynamic limit, a second order phase transition. 
 
We have performed numerics for small system sizes ($n\leq 7, N_c\leq 20$) and observed the signatures of such phase transition (Sec.\ \ref{sec:hysteresis}). In order to exclude a first order transition, we have investigated the hysteresis phenomenon, not finding any signature of it.  The quantitative comparison of the analytic, large $N_c$ results, with the numerics is plagued by large $N_c$ corrections (larger, the larger is $n$), but it is to a large extent satisfactory. 

The frustration generated by the entanglement monogamy is to some extent similar to what happens in frustrated spin systems which have (for low temperatures) a glassy phase. However, it is worth remembering that our model does not contain any quenched disorder. Taking this into account, a more fit analogy is with configurational glasses~\cite{marinari1,marinari2,borsari}.

Our result is of considerable significance for the physics of entanglement. Decreasing the temperature, the average of the potential of multipartite entanglement decreases towards its minimum. A smaller value of the potential is representative of a \emph{more entangled state}. The typical states at low temperature are sampled from a measure which is not $\cS_n$ symmetric, so the states which minimize the potential are necessarily concentrated on fewer spin configurations. This is at first thought counterintuitive. One might think that in order to get \emph{more} multipartite entanglement one should mix in more configurations and create a more uniform wave vector on the {binary hypercube $\mathbb{Z}_2^n$}. We show that the opposite is true. The {states} which maximize the multipartite entanglement have a certain degree of breaking of the $\cS_n$ symmetry.
 
Finally, there are other features of entanglement frustration that have not been analyzed in this article, that pertain to the ground states of some quantum many-body Hamiltonians \cite{MGI,DN,WVC,GAI,BOOE,Schnack}. As entanglement is a resource for quantum computation, it would be desirable to see if one can find a quantum {Hamiltonian} whose ground state is highly entangled. In this direction, recent developments have shown that some quantum many-body Hamiltonians \cite{Shor1,Shor2,Shor3} made from projectors \cite{Laumann1,Laumann2} do violate the common area law for entanglement in favor of volume or almost-volume law. These ground states are likely candidates for quantum certificates of difficult quantum computation problems.

\acknowledgements
This work was partially supported by PRIN 2010LLKJBX on ``Collective quantum phenomena: from strongly correlated systems to quantum simulators,'' and by the Italian National Group of Mathematical Physics (GNFM-INdAM, Progetto Giovani).  We acknowledge the $Bc^2 S$-RECAS farm (Universit\`a di Bari and INFN) for computational resources.


\begin{thebibliography}{99}

\bibitem{EPR}
A. Einstein, B. Podolsky, and N. Rosen, Phys. Rev. \textbf{47}, 777 (1935).

\bibitem{Schr}
E. Schr\"odinger, Proc. Camb. Phil. Soc. \textbf{31}, 553 (1935), ibid. \textbf{32}, 446 (1936).

\bibitem{woot}
W. K. Wootters, Quantum Inf. Comput. \textbf{1}, 27 (2001),

\bibitem{fazio}
L. Amico, R. Fazio, A. Osterloh, and V. Vedral, Rev. Mod. Phys. \textbf{80}, 517 (2008).

\bibitem{horo}
R. Horodecki, P. Horodecki, M. Horodecki, and K. Horodecki, Rev. Mod. Phys. \textbf{81}, 865 (2009).

\bibitem{nielsen_chuang}
M. A. Nielsen and I. L. Chuang, \textit{Quantum Computation and Quantum Information} (Cambridge University Press, Cambridge, 2000).

\bibitem{BZ}
I. Bengtsson and K. \.Zyczkowski, \textit{Geometry of Quantum States} (Cambridge University Press, Cambridge, 2006).


\bibitem{mw}
D. A. Meyer and N. R. Wallach, J. Math. Phys. \textbf{43}, 4273 (2002).


\bibitem{MMSZ}
V. I. ManÕko, G. Marmo, E. C. G. Sudarshan, and F. Zaccaria, J. Phys. A: Math. Gen. \textbf{35} ,7137 (2002)

\bibitem{MPV} 
M. Mezard, G. Parisi and M. A. Virasoro, \textit{Spin Glass Theory and Beyond} (World Scientific, Singapore, 1987) 

\bibitem{frustration}
P. Facchi, G. Florio, U. Marzolino, G. Parisi, S. Pascazio New J. Phys. \textbf{12}, 025015 (2010).


\bibitem{ckw}
V. Coffman, J. Kundu, and W. K. Wootters, Phys. Rev. A \textbf{61}, 052306 (2000).

\bibitem{verstraete}
T. J. Osborne and F. Verstraete, Phys. Rev. Lett. \textbf{96}, 220503 (2006).

\bibitem{adesso}
B. Regula, S. Di Martino, S. Lee, and G. Adesso, Phys. Rev. Lett. \textbf{113}, 110501 (2014) .



\bibitem{gourwallach}
G. Gour and N. R. Wallach, J. Math. Phys. \textbf{51}, 112201 (2010). 


\bibitem{HS00} 
A. Higuchi and A. Sudbery, Phys. Lett. A \textbf{273}, 213 (2000).

\bibitem{KNM}
V. M. Kendon, K. Nemoto, and W. J. Munro, J. Mod. Opt. \textbf{49}, 1709 (2002).

\bibitem{BSSB}
I. D. K. Brown, S. Stepney, A. Sudbery, and S. L. Braunstein, J. Phys. A \textbf{38}, 1119 (2005).
 
 
\bibitem{OS}
A. Osterloh and J. Siewert, Int. J. Quantum Inf. \textbf{4}, 531 (2006).
 
\bibitem{BPBZCP}
A. Borras, A. R. Plastino, J. Batle, C. Zander, M. Casas, and A. Plastino, J. Phys. A \textbf{40}, 13407 (2007).

\bibitem{BH07}
S. Brierley and A. Higuchi, J. Phys. A \textbf{40}, 8455 (2007).

\bibitem{scott}
A. J. Scott, Phys. Rev. A \textbf{69}, 052330 (2004).

\bibitem{mmes}
P. Facchi, G. Florio, G. Parisi, S. Pascazio, Phys. Rev. A \textbf{77}, 060304(R) (2008).

\bibitem{facchi_lincei}
P. Facchi,
Rend. Lincei Mat. Appl. \textbf{20}, 25 (2009).

\bibitem{AC} 
L. Arnaud and N. J. Cerf, Phys. Rev. A 87, 012319 (2013).

\bibitem{zycz}
D. Goyeneche and K. \.Zyczkowski, Phys. Rev. A \textbf{90}, 022316 (2014).

\bibitem{FFP}
P. Facchi, G. Florio, S. Pascazio, Phys. Rev. A \textbf{74}, 042331 (2006).

\bibitem{multipartite}
P. Facchi, G. Florio, U. Marzolino, G. Parisi, S. Pascazio
J. Phys. A: Math. Theor. \textbf{42}, 055304 (2009).

\bibitem{classical}
P. Facchi, G. Florio, U. Marzolino, G. Parisi, S. Pascazio
J. Phys. A: Math. Theor. \textbf{43}, 225303 (2010).

\bibitem{depasquale3}
A. De Pasquale, P. Facchi, V. Giovannetti, G. Parisi, S. Pascazio, and A. Scardicchio,
J. Phys. A: Math. Theor. \textbf{45}, 015308 (2012).

\bibitem{Facchi2} 
P. Facchi, U. Marzolino, G. Parisi, S. Pascazio, and A. Scardicchio, Phys. Rev. Lett. \textbf{101}, 050502 (2008).

\bibitem{depasquale2}
A. De Pasquale, P. Facchi, G. Parisi, S. Pascazio, and A. Scardicchio, Phys. Rev. A \textbf{81}, 052324 (2010).

\bibitem{Zyc2}
K. \.{Z}yczkowski,   H.-J. Sommers, 
J. Phys. A \textbf{34} 7111 (2001).

\bibitem{qcd}
G. 't Hooft, Nucl. Phys. B\textbf{75}, 461 (1974).

\bibitem{parisi}
E. Br\'{e}zin, C. Itzykson, G. Parisi, J. B. Zuber, Comm. Math. Phys. \textbf{59}, 35 (1978). 

\bibitem{jzj}
M. Moshe and J. Zinn-Justin, Phys. Rept. \textbf{385}, 69 (2003).

\bibitem{CKPR} L. F. Cugliandolo, J. Kurchan, G. Parisi, and F. Ritort, Phys.
Rev. Lett. 74, 1012 (1995).

\bibitem{Gaussian1}
P. Facchi, G. Florio, C. Lupo, S. Mancini, S. Pascazio,
Phys. Rev. A \textbf{80}, 062311 (2009).

\bibitem{anneal1}
S. Kirkpatrick, C. D. Gelatt, and M. P. Vecchi, Science \textbf{220}, 671 (1983).

\bibitem{anneal2}
D. P. Landau and K. Binder, \textit{A Guide to Monte Carlo Simulations in Statistical Physics} (Cambridge University Press, Cambridge, 2009).

\bibitem{marinari1}
E. Marinari, G. Parisi, and F. Ritort, J. Phys. A: Math. Gen \textbf{27}, 7615 (1994).

\bibitem{marinari2}
E. Marinari, G. Parisi, and F. Ritort, J. Phys. A: Math. Gen \textbf{27}, 7647 (1994).

\bibitem{borsari}
I. Borsari, M. Degli Esposti, S. Graffi, and F. Unguendoli, J. Phys. A: Math. Gen. \textbf{30}, 155 (1997). 



\bibitem{DN} 
C. M. Dawson and M. A. Nielsen, Phys. Rev. A \textbf{69}, 052316 (2004).

\bibitem{WVC} 
M. M. Wolf, F. Verstraete, and J. I. Cirac, Int. J. Quantum. Inform. \textbf{1}, 465 (2003).

\bibitem{BOOE} 
N. de Beaudrap, M. Ohliger, T. J. Osborne, and J. Eisert, Phys. Rev. Lett. \textbf{105}, 060504 (2010).

\bibitem{Schnack} 
J. Schnack, Dalton Trans. \textbf{39}, 4677 (2010).

\bibitem{GAI} 
S. M. Giampaolo, G. Adesso, and F. Illuminati, Phys. Rev. Lett. \textbf{104}, 207202 (2010).

\bibitem{MGI}
U. Marzolino, S. M. Giampaolo, and F. Illuminati, Phys. Rev. A \textbf{88}, 020301(R) (2013).

\bibitem{Shor1}
S. Bravyi, L. Caha, R. Movassagh, D. Nagaj, and P.W. Shor, Phys. Rev. Lett. \textbf{109}, 207202 (2012).

\bibitem{Shor2}
R. Movassagh, E. Farhi, J. Goldstone, D. Nagaj, T. J. Osborne, and P. W. Shor. Phys. Rev. A \textbf{82}, 012318 (2010).

\bibitem{Shor3}
R. Movassagh, and P.W. Shor.  arXiv preprint arXiv:1408.1657 (2014).

\bibitem{Laumann1}
C. R. Laumann, R. Moessner, A. Scardicchio, and S. L. Sondhi, in \textit{Modern Theories of Many-Particle Systems in Condensed Matter Physics}, pp. 295-332 (Springer Berlin Heidelberg, 2012).

\bibitem{Laumann2}
C. R. Laumann, R. Moessner, A. Scardicchio, and S. L. Sondhi. Quantum Information and Computation \textbf{10}, 1 (2010).

\end{thebibliography}
\end{document}